\documentclass[letterpaper,12pt,preprint]{aastex}

\usepackage{amssymb,amsmath,amsbsy}
\usepackage{booktabs}
\usepackage{bbold}
\usepackage{mathrsfs}
\usepackage{graphicx}
\usepackage[caption=false]{subfig}
\usepackage[backref,breaklinks,colorlinks,citecolor=blue]{hyperref}
\usepackage[yyyymmdd]{datetime}

\usepackage[normalem]{ulem}

\newcommand{\mean}[1]{\left< #1 \right>}
\newcommand{\msun}{\ensuremath{\mathrm{M}_\odot}}
\newcommand{\bs}[1]{\boldsymbol{#1}}

\newcommand{\inttime}{t_{\rm int}}
\newcommand{\periods}{\ensuremath{{\rm T}}}

\newcommand{\project}[1]{\textsl{#1}}
\newcommand{\superfreq}{\project{SuperFreq}}
\newcommand{\chchchanges}[1]{#1}

\usepackage{color}

\begin{document}

\title{Chaotic dispersal of tidal debris}
\author{Adrian M. Price-Whelan\altaffilmark{\colum,\adrn},
	    Kathryn V. Johnston\altaffilmark{\colum},
	    Monica Valluri\altaffilmark{\mich},
	    Sarah Pearson\altaffilmark{\colum},
	    Andreas H. W. K\"upper\altaffilmark{\colum,\hubble},
	    David W. Hogg\altaffilmark{\nyu,\cds,\mpia}}
%\date{\centering \today}

% Affiliations
\newcommand{\colum}{1}
\newcommand{\adrn}{2}
\newcommand{\mich}{3}
\newcommand{\hubble}{4}
\newcommand{\nyu}{5}
\newcommand{\cds}{6}
\newcommand{\mpia}{7}
\altaffiltext{\colum}{Department of Astronomy,
		              Columbia University,
		              550 W 120th St.,
		              New York, NY 10027, USA}
\altaffiltext{\adrn}{To whom correspondence should be addressed: adrn@astro.columbia.edu}
\altaffiltext{\mich}{Department of Astronomy,
			   University of Michigan,
			   Ann Arbor, MI 48109, USA}
\altaffiltext{\hubble}{Hubble fellow}
\altaffiltext{\nyu}{Center for Cosmology and Particle Physics,
                      Department of Physics, New York University,
                      4 Washington Place, New York, NY, 10003, USA}
\altaffiltext{\cds}{Center for Data Science,
                      New York University,
                      4 Washington Place, New York, NY, 10003, USA}
\altaffiltext{\mpia}{Max-Planck-Institut f\"ur Astronomie,
                     K\"onigstuhl 17, D-69117 Heidelberg, Germany}

\begin{abstract}

Several long, dynamically cold stellar streams have been observed around the Milky Way Galaxy, presumably formed from the tidal disruption of globular clusters.
In integrable potentials---where all orbits are regular---tidal debris phase-mixes close to the orbit of the progenitor system.
However, the Milky Way's dark matter halo is expected not to be fully integrable; an appreciable fraction of orbits will be chaotic.
This paper examines the influence of chaos on the phase-space morphology of cold tidal streams.
\chchchanges{Streams even in weakly chaotic regions look very different from those in regular regions.}
We find that streams can be sensitive to chaos on a much shorter time-scale than any standard prediction (from the Lyapunov or frequency-diffusion times).
For example, on a weakly chaotic orbit with a chaotic timescale predicted to be $>$1000 orbital periods ($>$1000 Gyr), the resulting stellar stream is, after just a few 10's of orbits, substantially more diffuse than any formed on a nearby but regular orbit.
We find that the enhanced diffusion of the stream stars can be understood by looking at the variance in orbital frequencies of orbit ensembles centered around the parent (progenitor) orbit.
Our results suggest that long, cold streams around our Galaxy must exist only on regular (or very nearly regular) orbits; they potentially provide a map of the regular regions of the Milky Way potential.
This suggests a promising new direction for the use of tidal streams to constrain the distribution of dark matter around our Galaxy.

\end{abstract}

\keywords{
}

\section{Introduction}\label{sec:introduction}

The dark-matter haloes of galaxies are thought to be triaxial in shape with axis ratios and alignments that depend on radius. However, despite suggestive evidence from a range of complimentary observational methods, this fundamental prediction from $\Lambda$CDM cosmology has not been conclusively verified. Around other galaxies, it is generally hard to measure the 3D mass profile because informative tracers are rare and the haloes are seen in projection. From the Earth's position within the Milky Way, our view of our own halo and proximity gives us a unique chance to directly measure the 6D positions of stars and model the shape of the mass distribution at large radii. The Milky Way halo has a low density of visible tracers, but luckily many of the halo stars are likely associated with debris stripped from stellar systems and thus contain additional information encoded by the formation mechanism.

As a satellite galaxy or globular cluster orbits within some larger system, mass is eroded by tidal forces from the host-galaxy potential. Along regular, mildly eccentric orbits, mass stripped from the progenitor has a small spread in orbital properties (e.g., energy, angular momentum). Once the debris has evolved far enough from the progenitor system that the self-gravity of the progenitor can be ignored (usually a fast process relative to the orbital time), the stars evolve essentially as an ensemble of test particle orbits in the potential of the host system. The debris remains coherent as a tidal stream if the phase-mixing time-scale is long: a small ensemble of regular orbits reaches a fully phase-mixed state in a timescale $\approx\sigma_\Omega^{-1}$, where $\sigma_\Omega$ is the dispersion in fundamental frequencies of the ensemble (tidal debris from a globular cluster typically has frequency spreads $\approx$0.1--1\%, so it can take hundreds to thousands of orbital periods to fully phase-mix). The ensemble spreads due to shearing from slight variations in their fundamental frequencies, which, for tidal streams, preferentially occurs along one dimension \citep{merritt96, helmi99}.

The morphological (density) evolution of the tidal debris therefore depends on the spread of orbital properties (e.g., actions or frequencies) of the debris and the orbit of the progenitor system, both of which are also determined by the shape and radial profile of the gravitational potential of the host galaxy. By modeling the observed phase-space density of stream stars along with the host galaxy potential it is hoped that we may infer the 3D mass distribution of the host. Many tidal streams are observed around the Milky Way, M31, and other nearby galaxies; the known streams span a large range of distances---from $\approx10$ to $100~{\rm kpc}$---and progenitor masses---from $\approx$10$^3$ to $\approx$10$^8~\msun$ in stellar mass---\citep[][]{ibata94,odenkirchen01,belokurov06,grillmair06a,grillmair06b,bonaca12}. There has been extensive work on developing methods to use data from these streams to measure properties of the Milky Way's dark matter halo. These methods span a range of complexity from orbit-fitting, to \emph{Streakline} \citep{kuepper12} or `particle-spray' models \citep{gibbons14}, to action-space density modeling \citep[e.g.,][]{sanders14, bovy14}, to N-body simulations \citep[e.g.,][]{law10}. All methods have been tested in some way on simulated observations of data and these tests typically demonstrate the recovery of parameters for analytic, static potential forms.

One example of stream modeling in a multi-component (static, analytic) potential was presented by \citet{pearson15}, who aimed to reproduce observations of the stellar stream density from the globular cluster Palomar 5 in a single oblate and single triaxial potential using \emph{Streakline} \citep{kuepper12} and N-body models. They used the observed number density of stream stars, a limited number of radial velocities for stream members, and the sky position, distance, radial velocity, and proper motion of the cluster itself to fit model streams to the data. In the oblate potential (a three-component bulge+disk+spherical halo potential), a thin model stream was easily found that reproduces the observed stellar density morphology of the stream. In the triaxial potential (the potential from \cite{law10}: a three component bulge+disk+triaxial halo fit to Sagittarius stream data), the model streams generically formed large, two-dimensional `fans' of debris near the ends, and no physically reasonable progenitor orbits could be found that reproduced the observed thinness and curvature of the stream given the observational constraints of the present-day position and velocity of the cluster. The result in \citet{pearson15} demonstrates that the morphology of a stream alone can be used to rule out a potential. With an understanding of the circumstances that lead to the differences in stream morphology, this could be a powerful tool for rejecting potentials from positional information alone.

The obvious difference between the two potentials considered by \citet{pearson15} is the extra symmetry of the oblate potential. It is well known that the number of degrees of freedom of a potential plays a critical role in determining the orbit structure of the potential; Hamiltonians with more than two degrees of freedom generically contain significant chaotic regions. \citet{pearson15} tested the stochasticity of the orbit of the progenitor that produced `fanned' debris by computing the Lyapunov exponent along this orbit but found that it is consistent with being regular over dynamically relevant timescales (many Hubble times). It has been shown previously that along some strongly chaotic orbits and in live cosmological haloes, tidal streams do form large, diffuse `fans' of debris \citep[e.g.,][]{fardal14, ngan15}, however it is unknown how the resultant properties of the debris (e.g., density or length of the stream) depend on the degree of stochasticity. The result from \citet{pearson15} suggests that even weak chaos (as measured by the Lyapunov exponent) may affect the density evolution and therefore observability of tidal streams. Understanding why this occurs and developing a measure to quantify the importance of this enhanced density evolution is a promising new direction to be explored further.

In this \emph{Article}, we study the effect of chaotic diffusion of the fundamental frequencies of individual orbits on the density evolution of tidal debris. We choose a simple, cosmologically motivated model for a triaxial potential, analyze the degree of chaos as computed from single-orbit diagnostics for grids of constant-energy orbits, and compare these results to measures of the density evolution of finite-volume ensembles of orbits (meant to mimic tidal debris). We find that even when the chaotic timescale is predicted to be long for a given orbit, chaos may manifest over much shorter times in small orbit ensembles through the chaotic diffusion of the constituent orbits. For a chaotic orbit, the frequency spectrum of the orbit evolves with time: for a small ensemble, the spread in frequencies is therefore time-dependent, which could enhance phase-mixing. This idea supports a reevaluation of the importance of chaos in galactic haloes and implies that the amount of chaos in a given potential may have significant consequences for the observability and survivability of thin, cold tidal streams.

This paper is organized as follows. We review relevant nonlinear dynamics in Section~\ref{sec:nldreview}. In Section~\ref{sec:methods}, we describe our choice of potential, method for numerical orbit integration, and introduce the chaos indicators used in this work. Our results are split into three subsections: in Section~\ref{sec:results1} we present iso-energy grids of orbits and discuss the orbit classes and chaotic timescales present in our potential; in Section~\ref{sec:results2} we study the density evolution of small ensembles of orbits around each orbit of the previous section; in Section~\ref{sec:results3} we describe the behavior of chaotic diffusion and use this to explain how chaos is relevant for tidal streams over short times. We discuss the implications of our results in Section~\ref{sec:discussion}, and conclude in Section~\ref{sec:conclusions}.

\section{Review of nonlinear dynamics}\label{sec:nldreview}

To explore the question of if and how chaos manifests in the density evolution of orbit ensembles over timescales much shorter than that predicted from generic chaos indicators, we must first understand the behavior of individual orbits in complex gravitational potentials and the orbital structures in the potential itself (i.e. the strength of resonances and chaos). An orbit in an $N$ degree of freedom (dof) Hamiltonian, $H$, is a set of $2N$ quasi-periodic time series,
\begin{equation}
(w_1(t),...,w_{2N}(t)) = (q_1(t),...,q_{N}(t),p_1(t),...,p_{N}(t)) \label{eq:coords}
\end{equation}
where $q_k$ and $p_k$ are conjugate coordinates in the sense that
\begin{align}
	\dot{p}_k &= -\frac{\partial H}{\partial q_k}\\
	\dot{q}_k &= \frac{\partial H}{\partial p_k}
\end{align}
for all $t$. If bounded, the motion in any component, $w_k(t)$, can be described as a Fourier sum,
\begin{equation}
	w_k(t) = \sum_j^\infty a_{kj} \, e^{i\,\omega_j\,t} \label{eq:fourier}
\end{equation}
where the $a_{kj}$ are complex amplitudes.

A \emph{regular orbit} is a set of such time series that can be transformed to a special set of canonical coordinates known as angle-action coordinates. In these coordinates, the position variables are angles, $\boldsymbol{\theta}$, that increase linearly with time with rates set by $N$ constant, fundamental frequencies, $\boldsymbol{\Omega} = (\Omega_1, ..., \Omega_N)$. The frequency of a Fourier component (the $\omega_k$ in Equation~\ref{eq:fourier}) for any individual component of motion are just linear, integer combinations of the fundamental frequencies, $\boldsymbol{\Omega}$---that is, for a regular orbit, any Fourier component frequency may be written
\begin{align}
	\omega_j &= \boldsymbol{n_j} \cdot \boldsymbol{\Omega} \label{eq:fourierfreq}
\end{align} %; \, (\boldsymbol{n}_k \in \mathbb{Z}^3)
where $\boldsymbol{n}_k$ is a vector of $N$ integers. The conjugate momentum coordinates---the actions,  $\boldsymbol{J}$---are constants of motion. Even stronger, the actions are isolating integrals and any pair are in involution such that
\begin{equation}
	[J_\alpha, J_\beta] = 0
\end{equation}
where $[\cdot,\cdot]$ is the Poisson bracket. This implies that for an $N$ dof system, a regular orbit has $N$ independent constants of motion and the motion is therefore restricted to an $N$-dimensional manifold embedded in the 2$N$ dimensional phase space. The topology of angle-action space is toroidal and any regular orbit in an $N$ dof Hamiltonian can be understood as motion on the surface of an $N$-torus. Each set of actions, $(J_1,...,J_N)$ (or frequencies), labels a torus, and regular orbits are sometimes referred to in terms of their orbital tori (see, e.g., Section 50 in \citealt{arnold78}, Section 10-5 in \citealt{goldstein80}, Section 3.5 in \citealt{binneytremaine}).

\subsection{Orbits in integrable potentials}

A Hamiltonian or potential is said to be \emph{globally integrable} when the number of isolating integrals of motion is equal to the number of degrees of freedom and a transformation to angle-action coordinates may be done globally---for example, the transformation to angle-action coordinates may be written as a function of arbitrary phase-space coordinates and the functional form is independent of phase-space position \citep[e.g.,][]{goldstein80}. The condition for global integrability is very restrictive and the likelihood that a Hamiltonian is globally integrable decreases as the number of degrees of freedom increase \citep[e.g.,][]{lichtenberg83} and as the potential becomes more complex (i.e. containing multiple components with different shapes). In a globally integrable potential, the Hamiltonian may be written solely in terms of the actions, $H = H(\boldsymbol{J})$. Galactic potentials are almost certainly not globally integrable but it is useful to understand the orbit structure in integrable systems before extending to more general potentials. \chchchanges{For example, there are four general classes of orbits in the `Perfect Ellipsoid' potential \citep[an integrable triaxial potential and special case of the St\"ackel potential; see, e.g.,][]{kuzmin73, deZeeuw85}: box, inner long-axis tube, outer long-axis tube, and short-axis tube orbits. Regular orbits in non-integrable triaxial potentials can typically still be identified with these classes. Tube orbits have a non-zero time-averaged angular momentum about either the long or short axis and therefore never pass through the center of the potential (hence are centrophobic). Box orbits instead have a zero time-averaged angular momentum and therefore have finite probability of passing through the center of the potential. These orbits are generally centrophilic, though some resonant box orbits are also centrophobic (e.g., banana, pretzel, fish).}

\begin{figure}[h]
\begin{center}
\includegraphics[width=\textwidth]{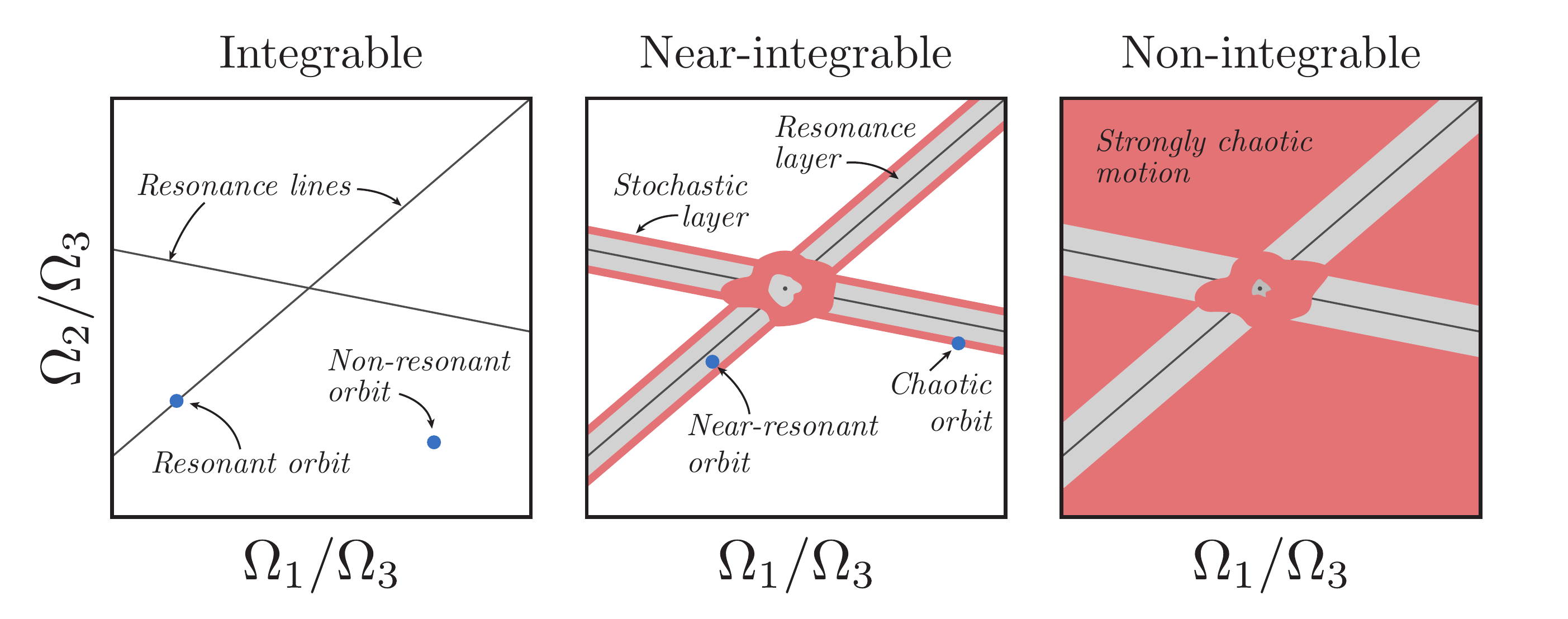}
\caption{An illustrative demonstration of the orbit structure for integrable (left), near-integrable (middle), and non-integrable (right) potentials in terms of the ratios of the fundamental frequencies. In an integrable potential, only resonant and non-resonant orbits exist, and all orbits are regular (resonances appear as lines in frequency-ratio-space). If the potential is perturbed mildly, resonance layers form around the resonances that host near-resonant orbits which behave like resonant orbits but have an additional frequency corresponding to libration about the resonance. Where resonances overlap or near separatrices, stochastic layers form and orbits will be chaotic. For more strongly perturbed potentials, many of the non-resonant orbits may become chaotic but resonance layers may grow and still remain regular. }
\label{fig:cartoons}
\end{center}
\end{figure}

The frequencies of a generic orbit are typically incommensurable---that is, $\bs{n} \cdot \bs{\Omega} \neq 0$ for any integer vector, $\bs{n}$, with reasonable magnitude.\footnote{A more precise condition is stated in terms of a diophantine condition, e.g., $|\bs{n} \cdot \boldsymbol{\Omega}| > \alpha \, |n|^{-\gamma}$ where $\alpha, \gamma>0$.} These non-resonant orbits uniformly cover the surface of an orbital torus. If instead there exists a relation of the form $\boldsymbol{n} \cdot \boldsymbol{\Omega} = 0$ the orbit is referred to as a resonant orbit. Resonant orbits are confined to a surface with lower dimensionality than the surface of an orbital torus, depending on the number of resonance relations obeyed; in a triaxial potential, orbits may obey either zero, one, or two resonance relations. We refer to orbits that obey a single resonance relation as \emph{uni-resonant} orbits, and if an additional resonance relation exists, \emph{bi-resonant}. Uni-resonant orbits in a triaxial potential are confined to a 2D surface, and bi-resonant orbits are closed 1D curves. The resonant structure of a potential---the relative importance of particular resonance integer vectors---is difficult to compute, but determines the global behavior of orbits in the potential. In plots of frequency ratios, the resonances appear as lines; Figure~\ref{fig:cartoons} (left panel) shows a cartoon portrait of a portion of frequency-space for an integrable potential with example resonance lines, non-resonant, and resonant orbits marked. For an integrable potential, all orbits are regular.

% Figure ??

\subsection{Orbits in near- and non-integrable potentials}

The orbit structure of near-integrable potentials can be understood by considering a Hamiltonian that is a small perturbation away from being globally integrable---that is, a Hamiltonian that may be written
\begin{equation}
	H(\boldsymbol{J}, \boldsymbol{\theta}) = H_0(\boldsymbol{J}) + \epsilon \, H_1(\boldsymbol{J}, \boldsymbol{\theta})
\end{equation}
where $H_0(\bs{J})$ represents an integrable Hamiltonian, and $\epsilon$ is a small parameter that determines the perturbation strength \citep[a description of perturbation theory applied to nonlinear Hamiltonians is given in][]{lichtenberg83}. When $0 < |\epsilon| \ll 1$, resonant surfaces become `thick' resonant layers, within which orbits are qualitatively similar to the parent resonant orbit \citep[e.g.,][]{merritt99}. These resonant layers are then generically surrounded by stochastic layers where chaotic motion occurs (in the vicinity of separatrices). Chaotic orbits are not bound to the surface of a torus and instead fill a (2N-1)-dimensional volume where a continuous set of tori would exist in an unperturbed Hamiltonian, $H_0$. While the frequency spectrum of sub-sections of a regular orbit are indistinguishable (i.e. measured over a finite interval of time), the frequency spectra of sub-sections of a chaotic orbit will evolve stochastically.

Figure~\ref{fig:cartoons} (middle panel) shows a cartoon of frequency space for a near-integrable potential (a small perturbation away from the potential in the left panel) assuming the resonances are stable. Much of the structure that was present in the integrable potential remains in the near-integrable case, but the differences are highlighted. Orbits in the resonant layers surrounding the resonances (grey) are near-resonant orbits that librate around the resonance and have finite thickness \citep[e.g.,][]{merritt99}. Chaotic orbits in the stochastic layer (red) behave erratically depending on the surrounding resonance structure. If the stochastic layer is small and the chaotic orbit is therefore confined, the orbit may behave nearly regular for long periods of time. If the resonance in the unperturbed Hamiltonian is unstable, all orbits associated with the resonance will be chaotic; unstable resonances form linear gaps in frequency-space (this is not shown in the cartoon but are seen in Figure~\ref{fig:logfreqs}). Thus, only some resonances in the perturbed system will retain the signature of a resonance.

For small values of $\epsilon$, many regular orbits survive and only small chaotic regions are introduced, especially in the vicinity of the intersection of two resonance lines \citep[commonly referred to as `resonance overlap'; see][]{chirikov60}. As the strength of the perturbation increases, eventually most tori associated with non-resonant motion will be destroyed. Figure~\ref{fig:cartoons} (right panel) qualitatively shows this phenomenon---the resonant and resonant-layer orbits may still be regular, but many or all of the non-resonant orbits will be chaotic. As the perturbation strength increases, eventually the uni- and bi-resonant tori are also destroyed---these are less susceptible to destruction from perturbations \cite[for a more quantitative illustration of this transition from integrability to global chaos, see Figure~9 in][]{valluri98}.

When $\epsilon$ is large, there is no general prediction for the resulting behavior, however it seems that more complicated and physically motivated triaxial potential models for galaxies follow the intuition gained from the small-perturbation picture described above, at least for certain parameter choices \citep[e.g.,][]{valluri98, merritt99}. We therefore expect a large number of regular orbits will survive---the so-called Kolmogorov–Arnold–Moser (KAM) tori---however the tori that survive will be separated by regions of chaotic motion. Any transformations to angle-action coordinates must be defined local to each resonance region due to the destruction of tori and chaotic motion which lead to discontinuous changes in orbital properties.

\subsection{The behavior of chaotic orbits in non-integrable potentials}\label{sec:behavior-chaotic}

Chaotic orbits have no orbital actions and only conserve energy (if the potential is time-independent). The orbits therefore do not have a single set of fundamental frequencies, but rather the frequencies that describe the character of motion evolve with time. In near-integrable potentials, the frequencies of consecutive sub-sections of a chaotic orbit diffuse both around resonance layers\footnote{Note that during this diffusion, the frequencies never exactly hit those of a KAM torus but evolve stochastically around these discrete, stable tori \citep[cf. Figure 2 in][]{laskar99}. (a sort of stochastic libration) and along the stochastic layers that surround the resonances (Arnold diffusion).} For weakly chaotic orbits, motion around a resonance layer can occur with a frequency close to the libration frequency of the nearby stable orbits in the resonance layer. Thus, if the resonance libration frequency is small, motion around a resonance can modulate the frequency spectrum of an orbit over an orbital time. However, the stochastic layers are often bounded in the direction orthogonal to the resonance by other stable, resonant regions so that the frequencies or actions can not change by large factors (unless there are other nearby resonances and overlapping stochastic layers, in which case the motion may be strongly chaotic).

The rate of diffusion along stochastic layers via Arnold diffusion depends on the local resonant structure and is hard to predict. This has been done analytically for simple potentials \citep[e.g.,][]{chirikov79}. For systems with $N>2$ dof, the stochastic layers connect and form an intricate network of stochasticity known as the Arnold web; an orbit that ergodically mixes over its energy hypersurface must traverse this web, though the timescales typical for this phenomenon are many thousands of orbital periods.

Arnold diffusion is not expected to be significant for most orbits over timescales relevant to galaxies (10s of orbits), however chaotic motion across resonances can occur over short times. If a stochastic trajectory is surrounded by regular orbits, it may be trapped around a regular parent orbit for long periods of time before escaping to another such semi-bounded region where it can become trapped around another parent orbit (a process by which it may eventually explore the whole Arnold web)---such orbits are commonly referred to as `sticky orbits.' Additionally, if the volume of the surrounded region in frequency space is comparable to the characteristic spread of frequencies in the tidal debris, this small-scale evolution will be important for tidal debris.

\subsection{Mixing of orbit ensembles}\label{sec:chaotic-mixing}

An ensemble of regular orbits (e.g., tidal debris) will phase-mix because of (small) differences in the fundamental frequencies of the orbits. The frequency distributions of thin streams are generally close to one-dimensional---that is, one eigenvalue of the distribution of frequencies for an orbit ensemble will be much larger than the others because of the local shape of the Hessian of the potential \citep{helmi99, sanders13a, bovy14}. Phase-mixing generically leads to power-law decay of the mean density of the ensembles: initially, the density decreases linearly in time because of the large, nearly one-dimensional spread in frequencies, then may proceed as $t^{-2}$ to $t^{-3}$ depending on relative sizes of the other eigenvalues of the frequency-space distribution \citep[e.g.,][]{helmi99, vogelsberger08}.

Generically, a small ensemble of chaotic orbits will lose coherence much faster than for regular orbits \cite[see, e.g.,][]{kandrup94, merritt96, kandrup03}, however this depends on the details of the resonant structure around the ensemble and the chaotic evolution of the individual orbits and is thus difficult to predict. For example, ensembles of orbits `stuck' between resonances may quickly spread to fill the allowed volume, but then the orbits must escape this confinement and diffuse through the Arnold web to reach a fully mixed state \citep{merritt96}. That is, while the orbit is stuck, the small-scale variations effectively cause an increase in the variance of the frequency distribution, which would enhance mixing of the debris in configuration-space. In this work, we investigate the consequences of short-time but small-scale frequency evolution and hypothesize that this may explain the enhanced density evolution of tidal debris around weakly chaotic regions where chaotic timescales are predicted to be long \citep[e.g.,][]{pearson15}. We then discuss how this would affect our understanding of the coherence and density evolution of tidal streams.

\section{Numerical methods}\label{sec:methods}

Our goal is to map the orbit structure of arbitrary (galactic) potentials, with an emphasis on identifying the chaotic orbits and understanding the evolution of these ensembles of orbits over short times. In particular, we aim to understand how this chaos-enhanced density evolution can affect tidal stream morphology. In this section, we describe the methods we will use to detect and quantify the strength of chaos for large grids of orbits.

\subsection{Potential choice}\label{sec:potential}

The density distributions within dark-matter haloes formed in cosmological N-body simulations are generically triaxial \citep[e.g.,][]{jing02, bett07, zemp09, veraciro11}. With the inclusion of baryonic physics and sub-grid prescriptions for energy input due to supernovae and other feedback mechanisms, the inner potential ($\lesssim$$0.1R_{\rm vir}$ for a $\approx$$10^{12}~\msun$ halo mass) typically becomes more spherical, though the magnitude of this reshaping depends on the particular merger history and star formation efficiency within a given halo and the mass and shape of the baryonic component \citep[e.g.,][though in Milky Way-like galaxies, baryonic disks will add non-sphericity to the total potential]{dubinski94,kazantzidis04, debattista08, bryan13, butsky15}. It is less clear what happens to the outer halo \citep[e.g.,][]{zemp11, valluri13}.

 \citet[][hereafter JS02]{jing02} found that a triaxial generalization of the NFW density profile \citep{navarro96} generates excellent fits to the density distributions within haloes in their high-resolution (dark-matter-only) N-body simulations, and they provide probability distributions for the axis ratios of a large sample of these haloes. JS02 find median axis ratios of $c/a \approx 0.55$ and $b/a \approx 0.77$ where $a$ is the major axis, $b$ the intermediate, and $c$ the minor axis.\footnote{Note that JS02 use the opposite notation so that $c$ is the major and $a$ is the minor axis.} These are largely consistent with findings from more recent simulations \citep[e.g.,][]{bett07, veraciro11, butsky15, zhu15} and consistent with constraints from weak lensing that place a lower limit on minor-to-major axis ratios of $c/a\gtrsim0.5$ \citep{vanuitert12}. JS02 find significant scatter in the distributions of concentration parameter, $c_e$, or scale radius (depending on choice of parametrization).

All of these parameters are specified in terms of the \emph{density}; for orbit analysis, we need to determine the form of the potential in terms of these parameters, which, in general, requires numerical integration of the density at each position of interest. For computational efficiency, many authors instead express the triaxiality in the form of the potential, but this can lead to unphysical situations where the density becomes negative. \citet{leesuto03} derive a perturbative expansion of the potential integral for a triaxial NFW density and show that the expansion is accurate even for modest axis ratios (e.g., the median values shown above).

\begin{figure}[h]
\centering
    \subfloat{
      \includegraphics[width=0.45\textwidth]{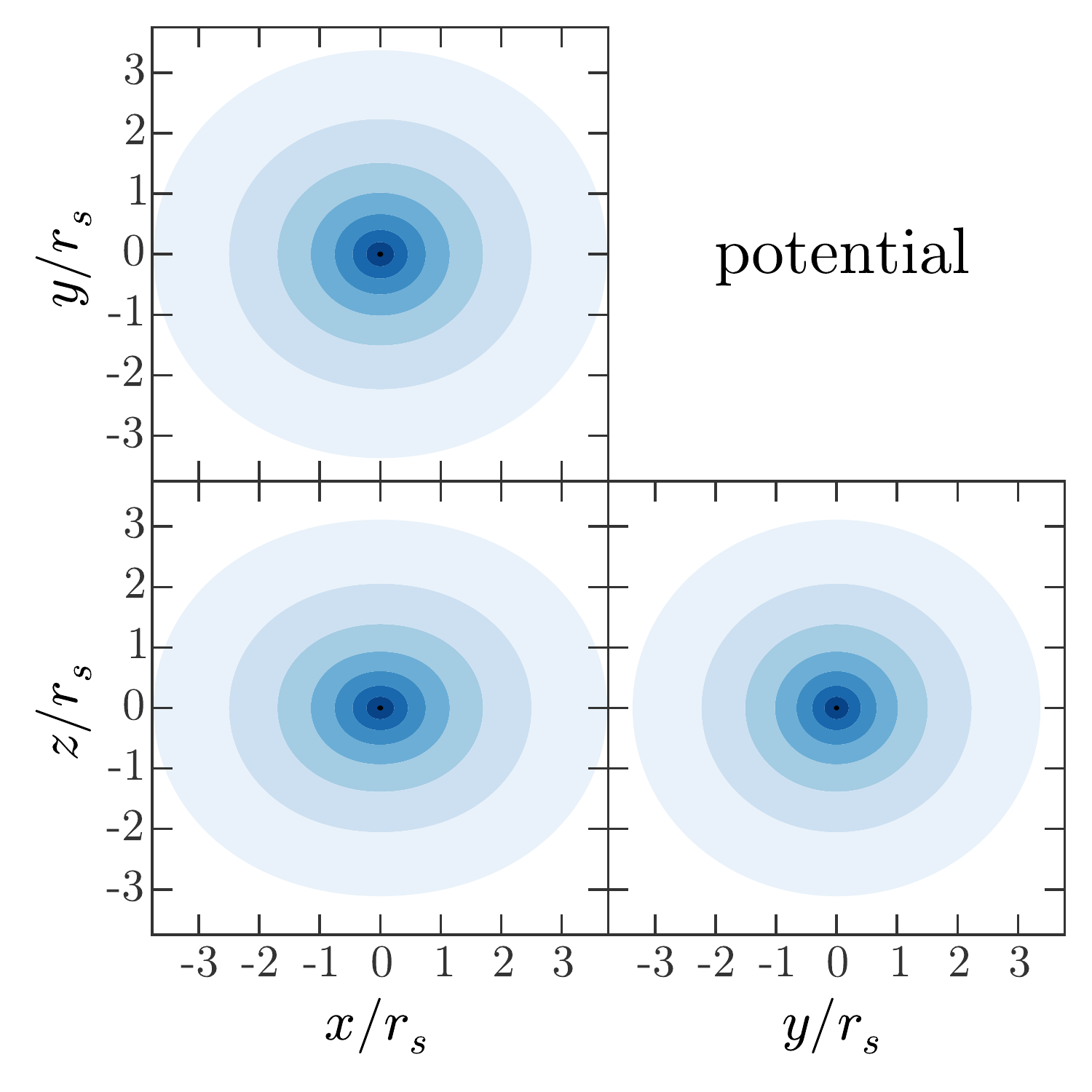}
    }
    \subfloat{
      \includegraphics[width=0.45\textwidth]{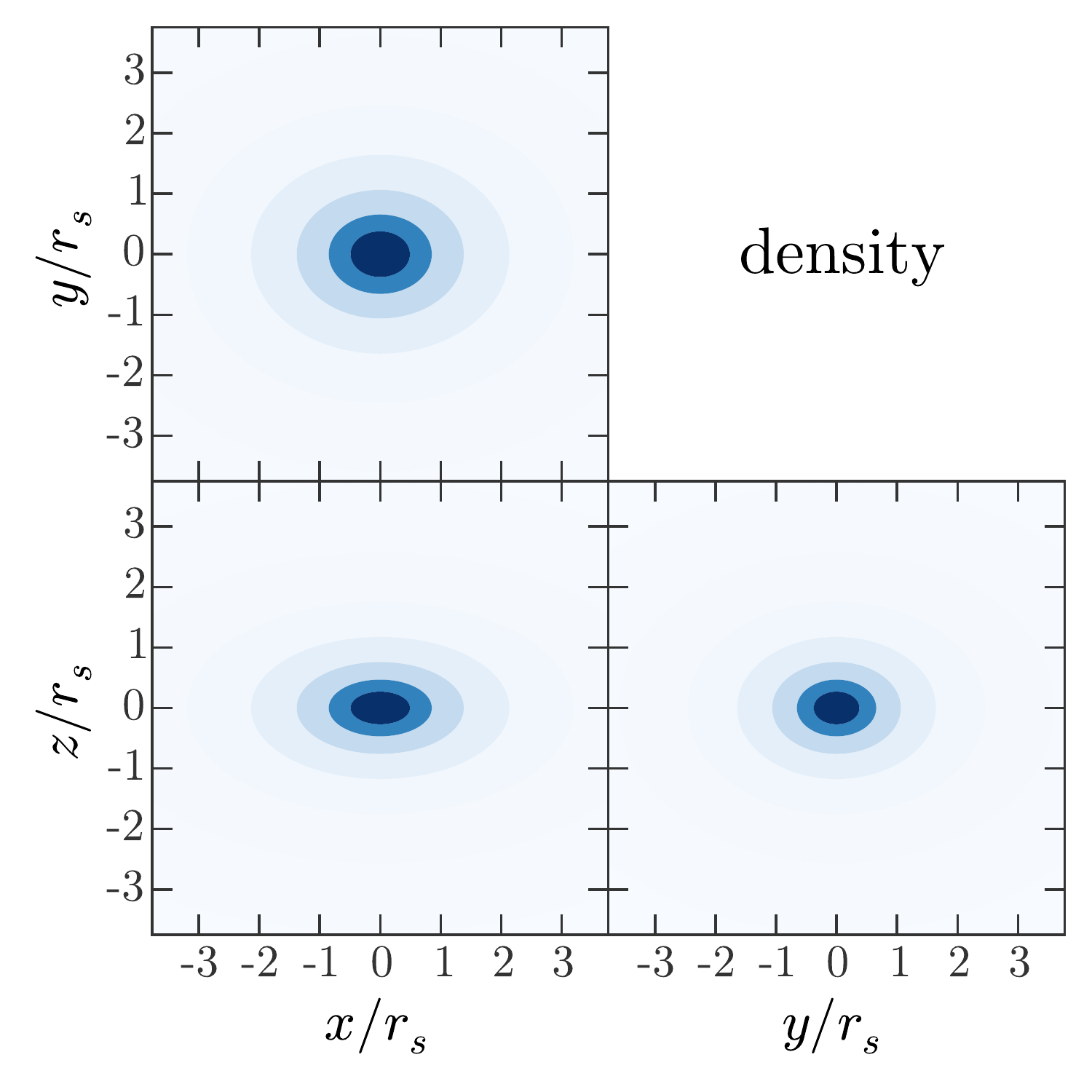}
    }
\caption{Equipotential contours (left) and isodensity contours (right) for the triaxial NFW potential considered in this work. For the potential plot there are eight contour levels evenly spaced and linear in the value of the potential. For the density plot there are eight contour levels logarithmically spaced from $10^4~\msun~{\rm kpc}^{-3}$ to $10^7~\msun~{\rm kpc}^{-3}$.}
\label{fig:potential}
\end{figure}

In this work, we use the triaxial potential expression from \citet{leesuto03}, parametrized in a slightly different manner. In terms of spherical coordinates\footnote{$(r,\phi,\theta)$ = (radius, azimuth, colatitude)} with the radius normalized by the scale radius, $u = r/r_s$
\begin{align}
	\Phi(u,\phi,\theta) &\approx \frac{v_c^2}{A}\left[F_1(u) + \frac{1}{2}(e_b^2 + e_c^2)F_2(u) + \frac{1}{2} [(e_b\sin\theta \sin\phi)^2 + (e_c\cos\theta)^2] F_3(u) \right]\label{eq:potential}\\
	A &= \left(\ln2 - \frac{1}{2}\right) + \left(\ln2-\frac{3}{4}\right) (e_b^2 + e_c^2)
\end{align}
where $e_b = \sqrt{1 - (b/a)^2}$, $e_c = \sqrt{1 - (c/a)^2}$, and $v_c$ is the circular velocity at the scale radius, $r_s$, for the spherical case. The functions $F_i(u)$ are given in the appendix of \cite{leesuto03}. We chose $r_s=20~{\rm kpc}$ and $v_c = 175~{\rm km}~{\rm s}^{-1}$ by taking the mean halo concentration for a ${\rm M}_{vir} \approx 10^{12}~\msun$ halo, $c_e\approx5$, from \cite{jing02} and by assuming $R_{vir}\approx200~{\rm kpc}$. Figure~\ref{fig:potential} shows equipotential contours of this potential, and Table~\ref{tbl:potential} summarizes the potential parameters.

\begin{table*}[ht]
\begin{center}
	\begin{tabular}{ c  c }
	         Parameter & Value\\\toprule
		$v_c$ & 175~km~s$^{-1}$\\
		$r_s$ & 20~kpc\\
		$a$ & 1\\
		$b$ & 0.77\\
		$c$ & 0.55\\
		\midrule
		$T_{abc}$ & 0.58\\
		\bottomrule
		\end{tabular}
	\caption{Summary of parameters for the triaxial NFW potential (Equation~\ref{eq:potential}) used in this work. Triaxiality is introduced in the density (rather than the potential) to ensure that the density is physical at all radii. Velocity scale, scale radius, and axis ratios are chosen to match the median halo parameters for a $M_{\rm vir} \approx 10^{12}~\msun$ halo from \citep{jing02}. The triaxiality parameter, $T_{abc} = \frac{a^2 - b^2}{a^2 - c^2}$, is also given. \label{tbl:potential}}
\end{center}
\end{table*}

This potential is a simple (and unrealistic) model for the total potential of a Milky-Way-like galaxy, however it represents a conservative choice for exploring the structure of orbits in the haloes of such galaxies. Realistic galactic potentials will have a significant component due to the disk and bulge, radially changing axis ratios or orientations \citep[e.g.,][]{romanowsky98, kazantzidis04,debattista08,veraciro11,butsky15}, significant substructure \citep{moore98,zemp09}, or time dependence \citep[either from bulk rotation, mass growth, mergers, etc.; see, e.g.,][]{bailin05}. We expect inclusion of any of these effects to increase the complexity of the resonant structure and influence of chaos (see Section~\ref{sec:discussion}).

\subsection{Orbit integration}\label{sec:integration}

We use the Dormand-Prince 8th-order Runge-Kutta scheme \citep{prince81} to integrate orbits in the above potential. Specifically, we use a \texttt{Python} wrapper over the \texttt{C} implementation by \cite{hairer93}. For all orbits we ensure that energy is conserved to $|\Delta E/E_0| \leq 10^{-8}$ by the end of integration, however most orbits conserve energy to a part in $\approx$$10^{-13}$. Unless otherwise specified the integration timesteps are chosen so that there are 512 steps per strongest orbital period component, but the integrator uses adaptive stepping between each main step in order to satisfy a specified tolerance (we set the absolute tolerance to $\approx$100 times machine precision, $\texttt{atol} = 10^{-13}$).

\subsection{Lyapunov exponents} \label{sec:lyap}

The most well-known method for assessing chaotic motion is to analyze the Lyapunov spectrum or maximum Lyapunov exponent (MLE) of an orbit \citep{lyapunov92}. The MLE measures the mean rate of divergence of two infinitesimally separated orbits and is only strictly defined in terms of a limit that goes to infinite time. Thus, we can never truly compute the MLE and it can take integration for many thousands of orbital periods to compute a converged numerical approximation of the MLE for a moderately chaotic orbit. In this work, we use the algorithm introduced by \cite{wolf85} for computing the MLE (for more a more detailed description of this algorithm, see Appendix~\ref{sec:lyapapdx}).

The MLE, $\lambda_{\rm max}$, is interpreted as a rate that quantifies the exponential divergence of infinitesimally close chaotic orbits. It is therefore useful to consider the corresponding $e$-folding time by inverting the rate,
\begin{equation}
	t_{\rm \lambda} = \frac{1}{\lambda_{\rm max}}.
\end{equation}
We will use this as the prediction from the Lyapunov exponent for the timescale over which chaos should be dynamically important for a given orbit.

\subsection{Frequency diffusion rate}\label{sec:naff}

Bounded, regular orbits in a triaxial potential have three fundamental frequencies, $\bs{\Omega}$, that determine the periodic behavior of motion. The motion in any canonical coordinate can therefore be decomposed as a Fourier sum (Equation~\ref{eq:fourier}) where the Fourier frequencies are linear, integer combinations of the fundamental frequencies (Equation~\ref{eq:fourierfreq}). \cite{laskar93} introduced a method for recovering the fundamental frequencies of an orbit that effectively uses fast-Fourier transforms (FFTs) of complex combinations of the motion (e.g., $x(t) + i v_x(t)$) to identify the frequencies. This method is referred to as `Numerical Approximation of Fundamental Frequencies' (NAFF) and has been used extensively in planetary dynamics \citep[e.g.,][]{laskar93b, laskar96} and galaxy dynamics \citep{papaphilippou98, valluri98}, especially in the study of orbits in triaxial systems. We have implemented and tested a version of this procedure in the \project{Python} programming language. Our implementation differs slightly from the original definition and from that used in \cite{valluri98}; we refer to this slight modification of the algorithm as \superfreq\ \citep{superfreq} and the code is open-source and publicly available on \project{GitHub}.\footnote{\url{https://github.com/adrn/SuperFreq}} For more details about the algorithm and differences with previous work, see \cite{laskar88, laskar93, papaphilippou96} and Appendix~\ref{sec:naffapdx}.

If an orbit is chaotic, the motion can no longer be expanded in terms of a single set of fundamental frequencies. For a weakly chaotic orbit, the orbit may appear consistently periodic over long windows of time. \superfreq\ will pick out a set of frequencies for chaotic orbits that correspond to the largest peaks in the power spectrum of the orbits, however these peaks will change location and amplitude with time. For more strongly chaotic orbits, the power spectrum will be quite noisy and the peak frequencies may change erratically when comparing two consecutive sections of orbit. The frequencies picked out by \superfreq\ for such orbits will therefore represent the average periodic nature of the orbit over a given integration window. Following previous work, we define the fractional frequency diffusion rate, $\mathcal{R}$, in the $k$th fundamental frequency as
\begin{equation}
	\mathcal{R}_k = \frac{\Omega_{k}^{(2)} - \Omega_{k}^{(1)}}{\Omega_{k}^{(1)} \, \Delta t} \label{eq:fdrate}
\end{equation}
where the upper index refers to the two consecutive sections of orbit and $\Delta t$ is the length of each integration window \citep{laskar93, valluri98, valluri12}. By inverting this rate, we can compute the timescale over which we expect order-unity changes to the fundamental frequencies: the \emph{frequency diffusion time} is defined as
\begin{equation}
	t_\Omega = \, (\max_{a_k} \, \mathcal{R}_k)^{-1} \label{eq:fdtime}
\end{equation}
where the maximum is taken with respect to the corresponding amplitudes, $a_k$, of the fundamental frequency components \citep[see][]{valluri12}.

For a small ensemble of orbits (e.g., tidal debris), a more relevant timescale is the time over which the change in frequencies for a single orbit is comparable to the spread of frequencies in the ensemble. We can estimate this timescale by multiplying the frequency diffusion time by a factor equal to the fractional spread in frequencies of the debris. For example, a globular cluster typically has $\approx$0.1\% spreads in fundamental frequencies, so by multiplying the frequency diffusion time by $f = 10^{-3}$, we can estimate the time (in number of orbital periods) over which we expect the frequencies to evolve by this amount.

% =====================================================================
%	Results
%
\section{Results}

In Section~\ref{sec:results1}, we generate grids of orbits in the potential described above to map the orbital structure of the potential. We classify each orbit in terms of the strength of chaos along the orbit as computed using the Lyapnov and frequency diffusion times of Section~\ref{sec:methods}. With this initial classification, in Section~\ref{sec:results2} we follow the evolution of ensembles of trajectories generated around each orbit in the initial grid. We find that the configuration-space density of ensembles around weakly chaotic orbits evolve faster (e.g., in mean density) than expected given the timescales over which chaos is computed to be relevant for the parent orbits. In Section~\ref{sec:results3}, we explain this phenomenon in the context of how chaotic diffusion occurs (e.g., Section~\ref{sec:behavior-chaotic}).

\subsection{Part I: Lyapunov and frequency diffusion times}\label{sec:results1}

\begin{figure}[h]%[p]
\centering
    \subfloat{
      \includegraphics[width=0.45\textwidth]{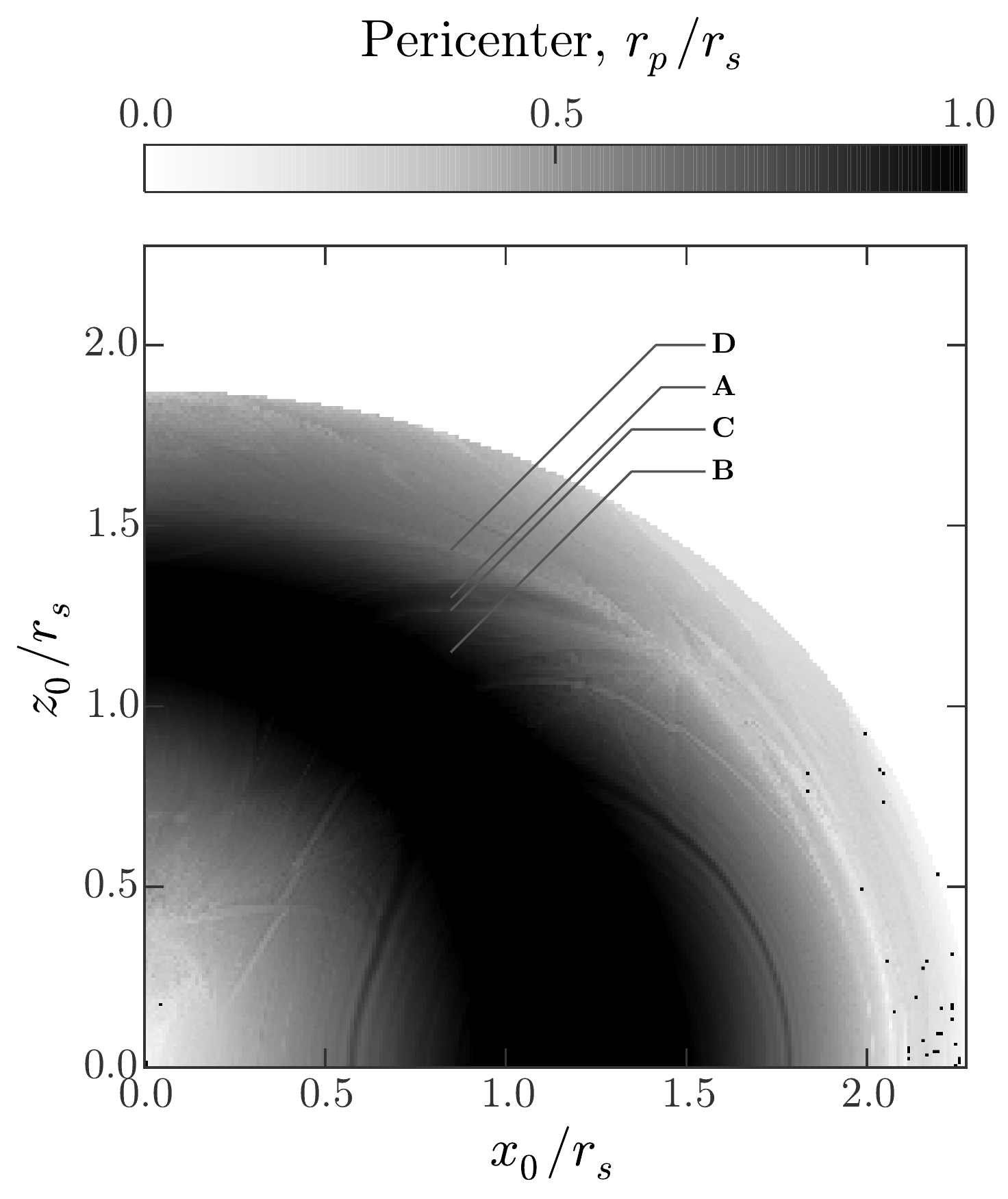}
    }
    \subfloat{
      \includegraphics[width=0.45\textwidth]{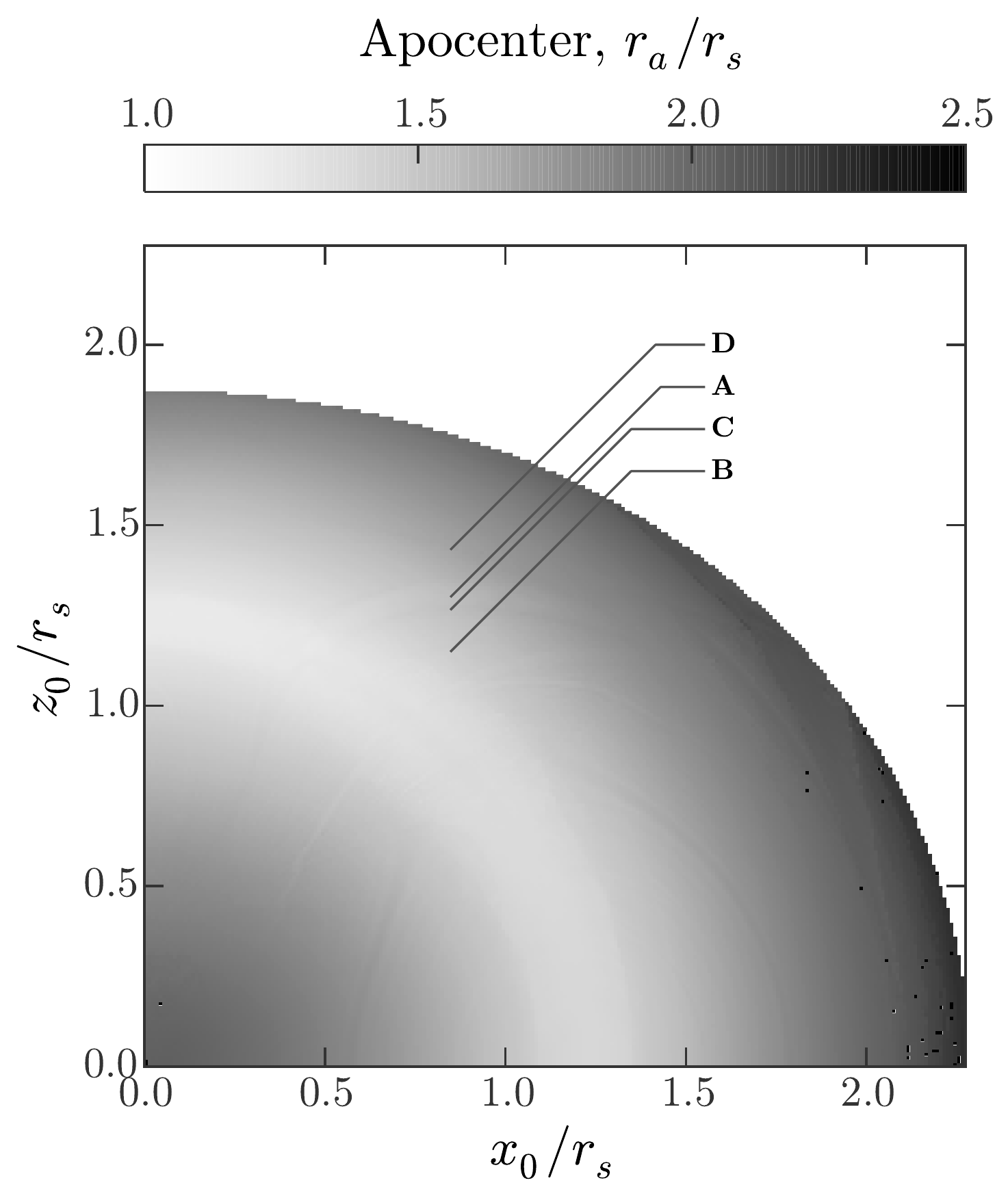}
    }
\caption{ A grid of isoenergy orbits initialized on the $xz$ plane. All distances are normalized by the potential scale radius. Each pixel in these panels represents a single orbit, and the shading of each pixel corresponds to the median pericentric distance (left) or median apocentric distance (right) computed over 64 orbital periods. The central black band in the left panel and white band in the right panel are tube orbits with apocenter-to-pericenter ratios close to one. The four arrows are explained in Section~\ref{sec:results2} and the caption of Figure~\ref{fig:lyapmap}.}
\label{fig:apoper}
\end{figure}

We generate an isoenergy grid of initial conditions along the $xz$ ($y=0$) plane\footnote{$x$ is the major and $z$ the minor axis.} with energy (per unit mass), $E$, chosen to span a range of distances comparable to the scale radius of the potential ($E=-(397.2~{\rm km~s}^{-1})^2$ in physical units; see Table~\ref{tbl:potential}). We fix $v_x = v_z = 0$, and compute $v_y$ from the energy. This grid generates all of the major orbit classes (short-axis tubes, inner long-axis tubes, outer long-axis tubes, stochastic intermediate-axis, and box orbits). The most numerous orbits in this grid are the short-axis and long-axis tubes that circulate about either the minor or major axis. Thin tidal streams may preferentially form along tube orbits rather than box orbits because of the faster disruption and debris diffusion expected for stellar systems on radially plunging orbits. Figure~\ref{fig:apoper} shows the grid of initial conditions in the $xz$ plane---each pixel in this grid represents an orbit, and the pixels are colored by the median pericentric distance (left) and median apocentric distance (right) over an integration time of 64 orbital periods. From here onwards, all lengths are given in units of scale radii, $r_s$, velocities in units of circular velocity, $v_c$, (see Table~\ref{tbl:potential}) and times in units of orbital periods, $\periods$.

% Figure ??

\begin{figure}[!h]%[p]
\begin{center}
\includegraphics[width=0.8\textwidth, trim={0 0 0 0}]{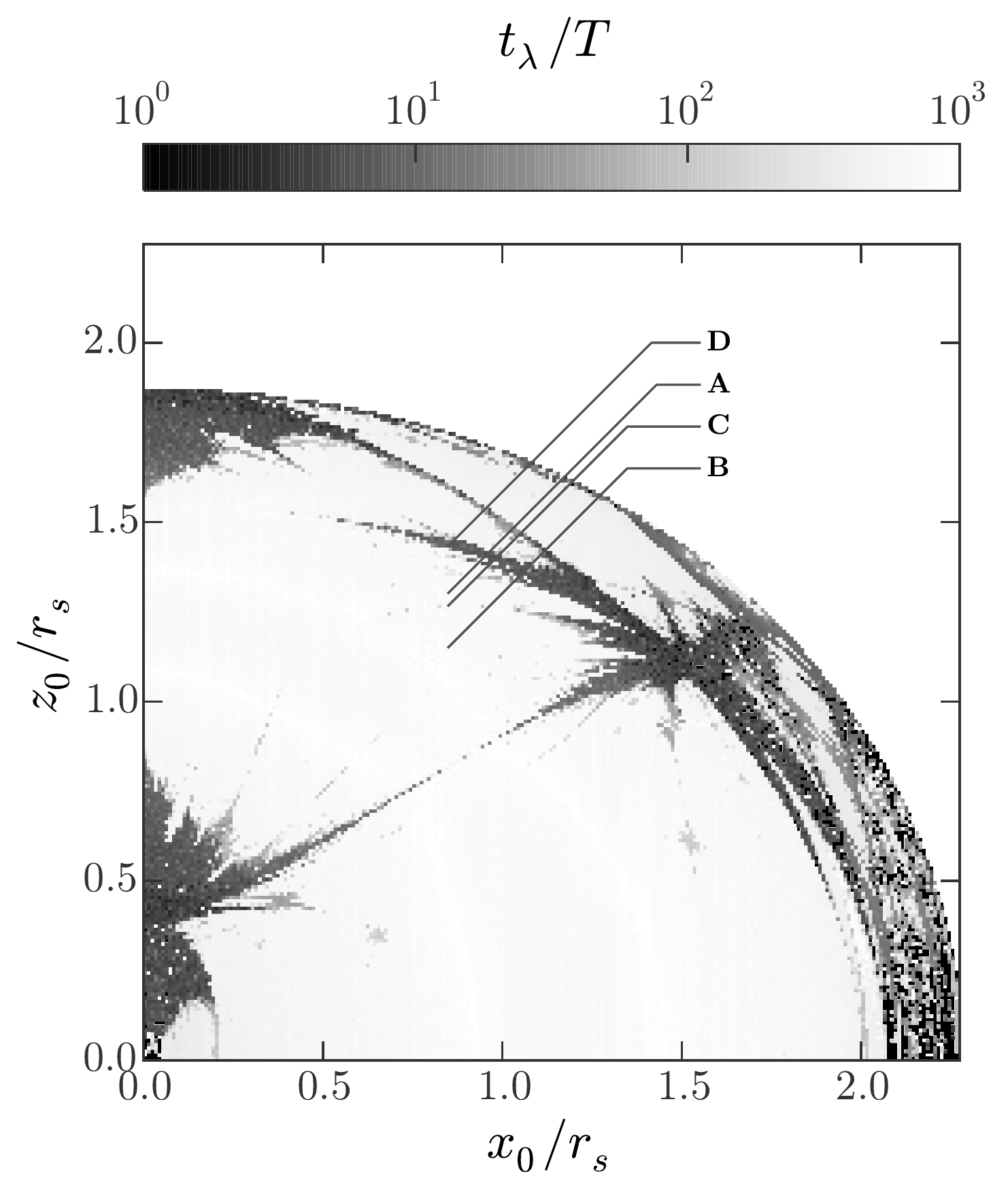}
\caption{ The same grid of orbits as shown in Figure~\ref{fig:apoper}, but now each pixel is colored by the logarithm of the Lyapunov time (in units of orbital periods). Orbits are integrated for a total of 10000 orbital periods. Chaotic orbits with $t_\lambda/\periods \gtrsim 700$ appear regular because the integration time for each orbit is insufficient to resolve weak chaos. Four orbits are pointed out with arrows---from top to bottom, these are the strongly chaotic (D), near-resonant (A), weakly chaotic (C), and non-resonant (B) orbits of Table~\ref{tbl:orbit-info} (see Section~\ref{sec:results2}).} \label{fig:lyapmap}
\end{center}
\end{figure}

We integrate all orbits in the grid for 10000 orbital periods and use the method described in Section~\ref{sec:lyap} to compute the MLEs. Figure~\ref{fig:lyapmap} again shows the grid of initial conditions, but now the color corresponds to the logarithm of the inverse of the MLE (the Lyapunov time). The darker pixels have shorter Lyapunov times and are more chaotic. Because of the fixed the integration time, the MLE cannot detect weak chaos and the majority of orbits appear to be regular because they have exceedingly long Lyapunov times (all white points have $(t_\lambda/\periods) \gtrsim 1000$).

% Figure ??

\begin{figure}[!h]%[p]
\begin{center}
\includegraphics[width=0.8\textwidth, trim={0 0 0 0}]{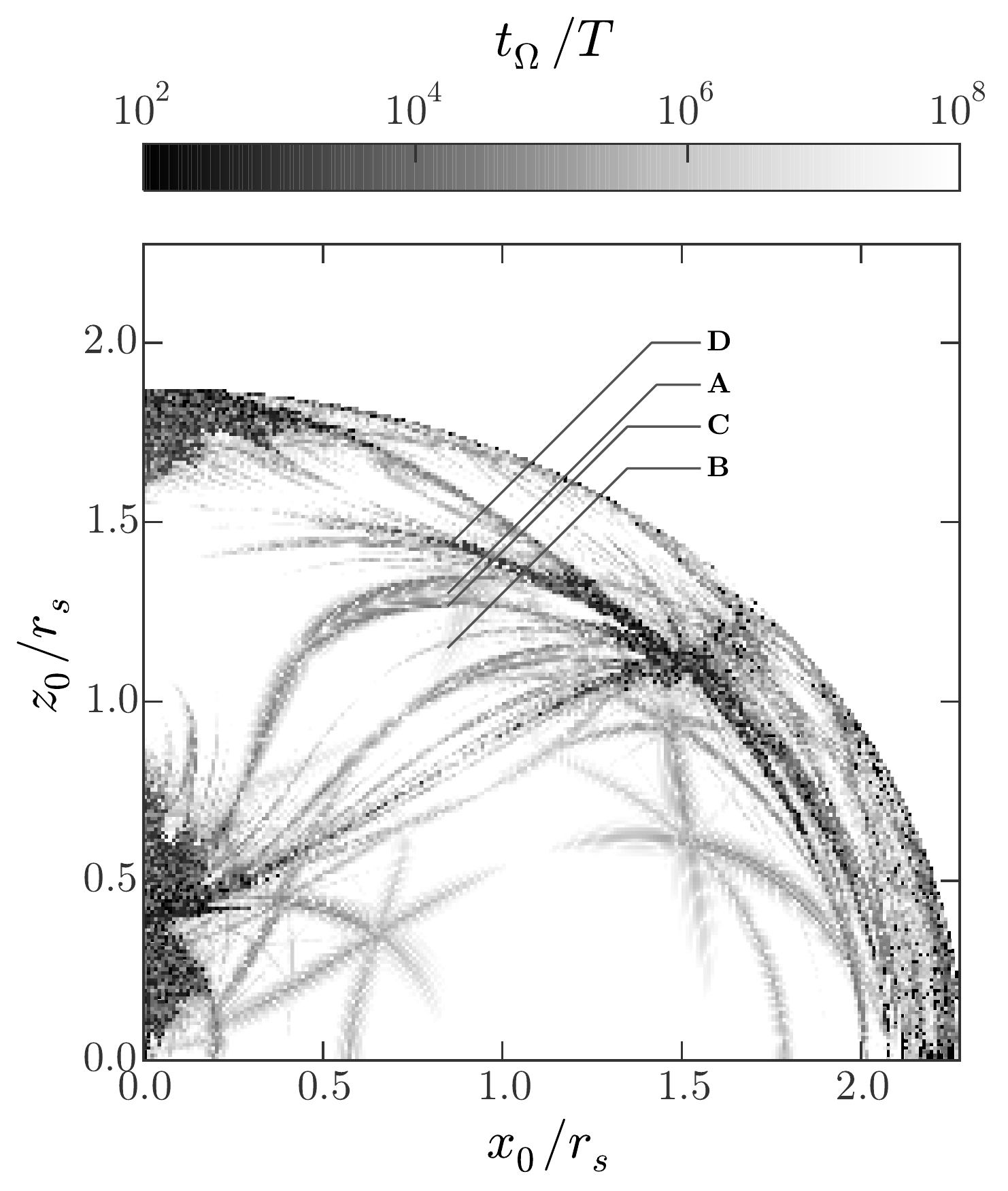}
\caption{The same grid of orbits as shown in Figure~\ref{fig:apoper}, but now each pixel is colored by the logarithm of the frequency diffusion time (again in units of orbital periods). The frequencies are computed in two consecutive windows, each of which has length equal to 40 orbital periods. The frequencies are measured precisely so that small changes in the frequencies can be detected over just $\approx$10s of orbits. Four orbits are pointed out with arrows---from top to bottom, these are the strongly chaotic (D), near-resonant (A), weakly chaotic (C), and non-resonant (B) orbits of Table~\ref{tbl:orbit-info} (see Section~\ref{sec:results2}).} \label{fig:freqdiff}
\end{center}
\end{figure}

For each orbit, we also separately integrate for 256 orbital periods and use \superfreq\ to compute the fundamental frequencies for the two consecutive sections of 128 orbital periods. We have chosen this window size so that we recover the frequencies for regular orbits with fractional error $\approx10^{-8}$ \citep[we estimate the error in frequency recovery using the method described in][]{laskar93}. With this integration window we are able to successfully recover frequencies for $>$99.9\% of the orbits \superfreq. Figure~\ref{fig:freqdiff} shows the same grid of initial conditions as in Figure~\ref{fig:lyapmap}, but now the greyscale intensity is set by the logarithm of the frequency diffusion time. The darker pixels have shorter frequency diffusion times and are more chaotic. This map reveals the intersection of this particular energy hypersurface with the rich structure of resonant surfaces present in this potential and highlights the accuracy of \superfreq\---weak chaos (grey) is detectable over much shorter integration periods using frequency analysis, compared to the many tens of thousands of orbits it would take to detect such features with the maximum Lyapunov exponent. The tube orbits in this potential are mostly regular or only mildly chaotic---the largest regular regions are associated with the short-axis and long-axis tube orbits---however islands of stronger chaos do appear, especially at the intersections of resonances where resonance overlap occurs.

% Figure ??

The strongest chaotic regions (black) appear in both of the above grids (Figure~\ref{fig:lyapmap} and Figure~\ref{fig:freqdiff}). Some of the weakly chaotic unstable resonances do appear in the Lyapunov time map---for example, near $(x_0,z_0) / r_s \approx(1.5,0.6)$ and $(x_0,z_0) /r_s \approx(0.6,0.4)$ where there is a slight hint of weak chaos (grey) in Figure~\ref{fig:lyapmap}. The details of the resonant structure is revealed in the frequency diffusion map from integrations of just 256 orbital periods. While there is rich structure and a significant number of weakly chaotic orbits, the majority of the orbits have estimated chaotic timescales corresponding to thousands of orbital periods and are thus not expected to be relevant for tidal stream evolution. In the next section, we analyze the density evolution of finite-volume ensembles of orbits around each orbit in the above grids in order to compare the effect of ordinary phase-mixing of tidal debris with potentially enhanced mixing due to chaos. We then compare the density evolution of the ensembles to the single-orbit chaos indicators computed in this section.

\subsection{Part II: Ensemble properties and mixing} \label{sec:results2}

The Lyapunov and frequency diffusion times measure the timescales over which chaos is relevant for a given orbit---that is, these quantities are measures of how infinitesimal deviations will diverge on average from some parent orbit, or of how long it takes for the frequencies of a single orbit to change by some amount. Tidal debris is disrupted from progenitor systems with finite spreads in orbital properties (e.g., energy). For a disrupting, globular-cluster-scale progenitor, the typical energy or frequency-space dispersion of the debris is 0.1--1\% of the progenitor orbital energy \citep[assuming masses of $10^4$--$10^6$~\msun;][]{johnston98}, but for a dwarf-galaxy-scale progenitor, the dispersion can be $\sim$10\%. In this section, we ask whether  the Lyapunov or frequency diffusion time predict the timescale over which a finite phase-space volume (e.g., tidal debris) stays coherent.

It is computationally intractable to run full N-body simulations for the large grid of orbital initial conditions of the previous section and we therefore take a simplified approach for studying how finite-volume debris spreads along each of these orbits. We instead consider small ensembles of particles meant to represent debris disrupted from a single tidal disruption event. For a given set of orbital initial conditions---the `parent' orbit---we integrate for 128 orbital periods, find the phase-space position of the minimum pericenter over this time, initialize a small ensemble of test particle orbits around this position, and integrate the orbits of all test particles for some integration time. We are interested in the degree to which chaos enhances the mixing rate of orbit ensembles and we therefore want to isolate out the effect of ordinary phase-mixing along regular orbits. We set the physical scale of the ensemble by the tidal radius in position and the velocity scale in velocity; the initial spread in fundamental frequencies will scale as $(m/M)^{1/3}$ like the tidal radius and velocity scale \citep[e.g.,][]{johnston98, apw14}. If $(\bs{x}_0,\bs{v}_0)$ are a set of initial conditions at pericenter for a parent orbit, then $\delta x_{k,i}$ and $\delta v_{k,i}$ are the $k$th components ($k=1,2,3$) of the position and velocity deviation vectors for the $i$th particle and the magnitude of the offsets are assumed to be Normally distributed away from the parent orbit:
\begin{align}
	\delta x_{k,i} &\sim \mathcal{N}(0, r_{\rm tide}/\sqrt{3})\\
	\delta v_{k,i} &\sim \mathcal{N}(0, \sigma_v/\sqrt{3})\\
	r_{\rm tide} &= \|\bs{x}_0\| \left(\frac{m}{M(<\|\bs{x}_0\|)}\right)^{1/3} \\
	\sigma_v &= \|\bs{v}_0\|\left(\frac{m}{M(<\|\bs{x}_0\|)}\right)^{1/3}
\end{align}
where $M(<r)$ is the mass enclosed of the host potential within radius $r$, $m$ is the mass scale of the `progenitor,' and $\|\cdot \|$ is the Euclidean norm. We take $m=10^4~\msun$ to represent globular-cluster-like progenitors, and use the spherically-averaged enclosed mass of the host potential to estimate the above debris scales. % we have verified this using regular orbits in a logarithmic potential...

\begin{figure}[h]%[p]
\begin{center}
\includegraphics[width=0.8\textwidth]{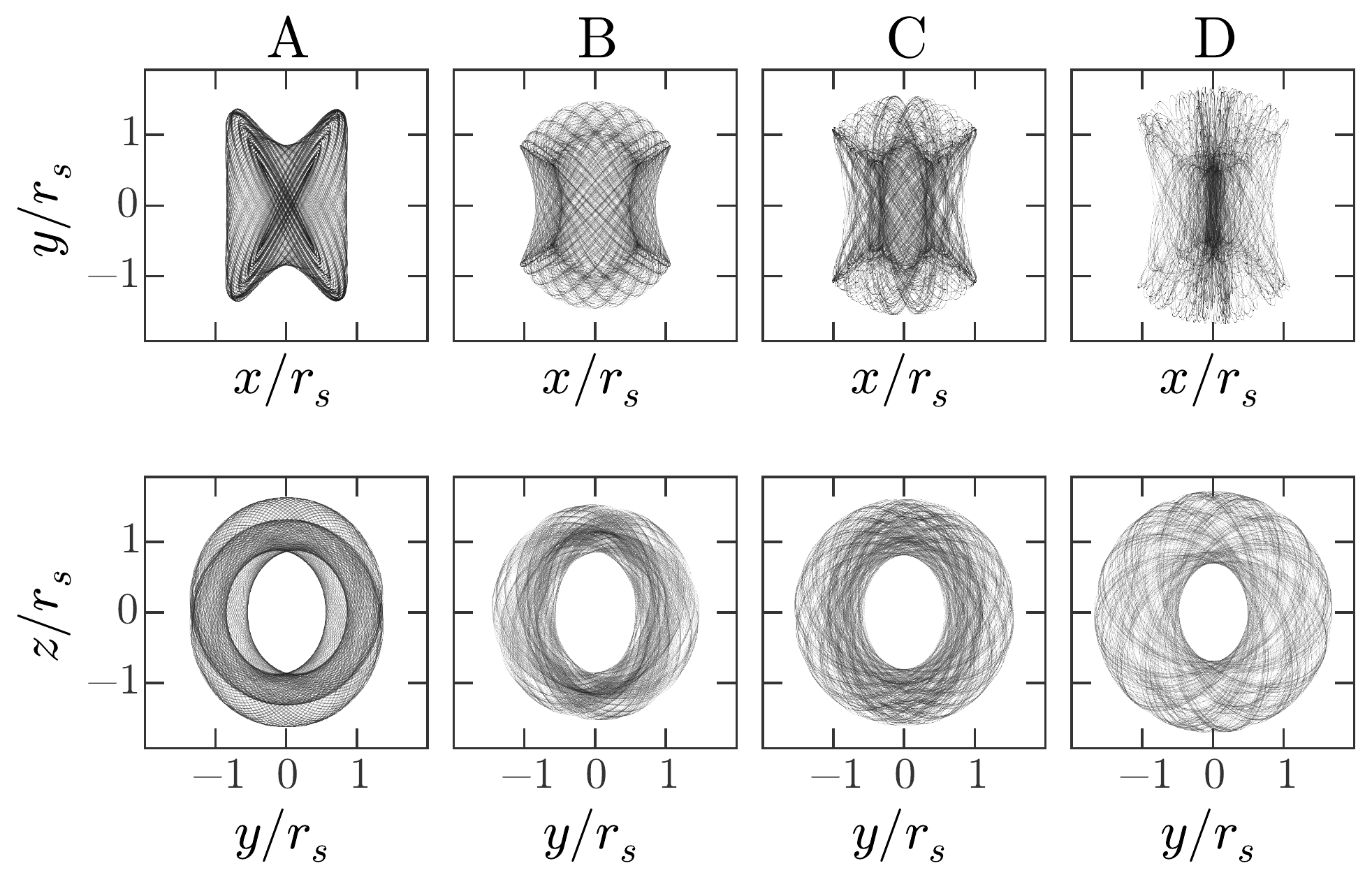}
\caption{Four representative orbits chosen from the orbit grid (e.g., Figure~\ref{fig:apoper}): a regular, near-resonant orbit (A), a regular, non-resonant orbit (B), a weakly chaotic orbit (C), and a more strongly chaotic orbit (D). All orbits are long-axis tubes with similar pericenters and apocenters. Orbits in this Figure were integrated for 1024 orbital periods and are shown in projection.}
\label{fig:orbits}
\end{center}
\end{figure}

\begin{figure}[h]%[p]
\begin{center}
\includegraphics[width=0.8\textwidth]{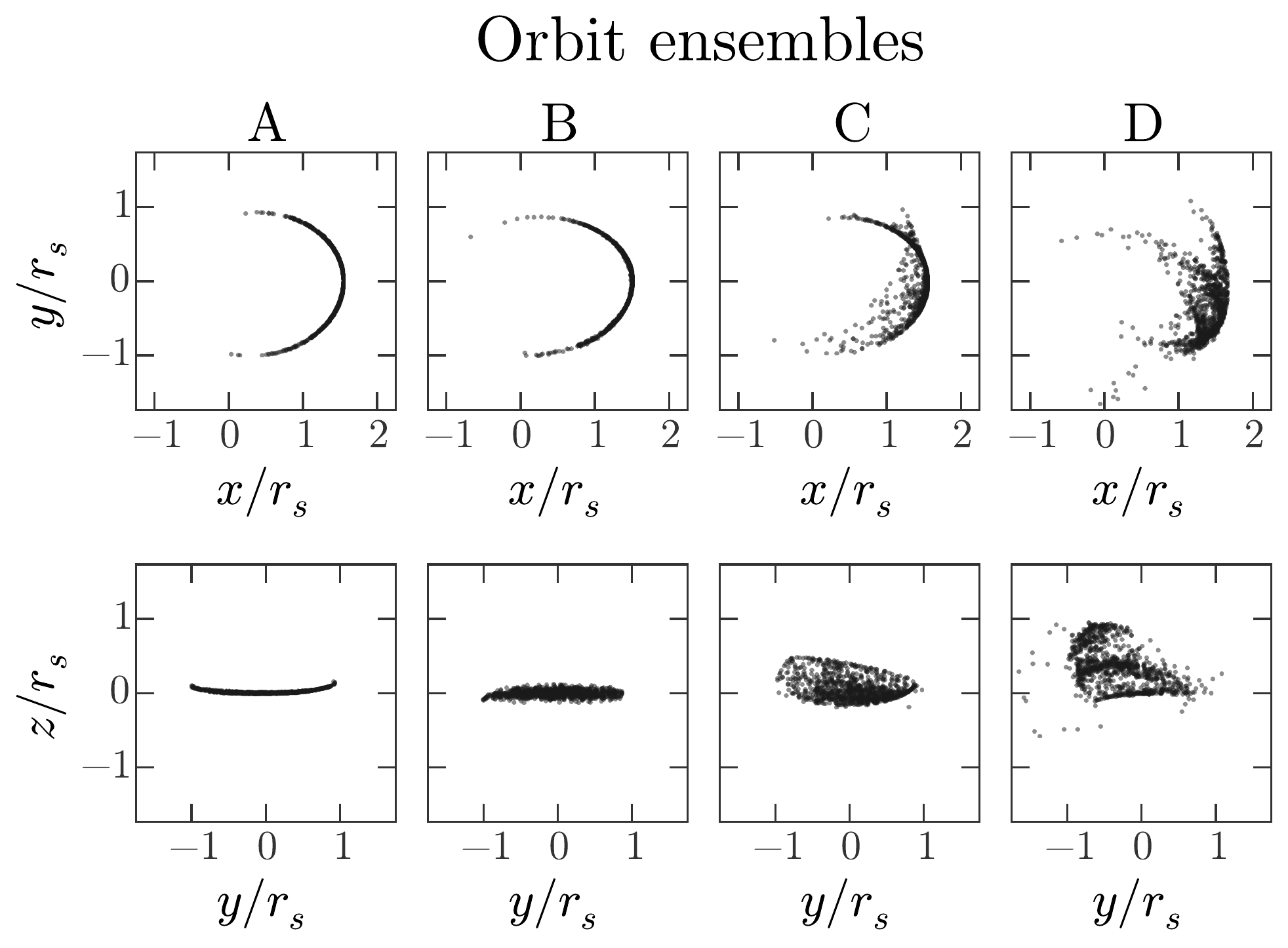}
\caption{Final particle positions after integrating unbound, globular-cluster-sized ensembles of orbits generated around each of the three N-body orbits (Figure~\ref{fig:nbodysims}). The ensembles each contain 1024 particles and are initialized at pericenter. The particle positions qualitatively match the morphology of the `oldest' debris from the corresponding N-body simulations (Figure~\ref{fig:nbodysims}). }
\label{fig:ensembles}
\end{center}
\end{figure}

We start by considering four particular orbits chosen from the orbit grid of Section~\ref{sec:results1}: a regular, near-resonant orbit (A), a regular, non-resonant orbit (B), a weakly chaotic orbit (C), and a more strongly chaotic orbit (D). The orbits were chosen to be close on the orbit grid so that their orbital properties (e.g., apocenter, pericenter) are similar, but have different frequency diffusion times; all four orbits are long-axis tube orbits. Figure~\ref{fig:orbits} shows the orbits in projection over an integration period of 1024 orbital periods. The initial conditions and chaos diagnostics for each orbit are listed in Table~\ref{tbl:orbit-info}. Figure~\ref{fig:ensembles} shows final positions of test-particle ensembles initialized around the four orbits described above and integrated for 64 orbital periods. A thin stream forms on the near-resonant orbit (left column), a thin---but more two-dimensional---stream forms on the non-resonant orbit (middle-left), a more diffuse stream forms on the weakly chaotic orbit with a slightly `fanned' morphology (middle-right), and a two-dimensional, `fanned' stream on the more strongly chaotic orbit (right). Given the long Lyapunov and frequency diffusion times of the weakly chaotic orbit ($t_\lambda/\periods > 900$, $t_\Omega/\periods \approx 2\times10^5$), it is surprising that the density evolution of the ensemble on this orbit appears to be more diffuse than the regular orbit stream.

% Figure ??

% Figure ??

\begin{figure}[h]%[p]
\begin{center}
\includegraphics[width=0.8\textwidth]{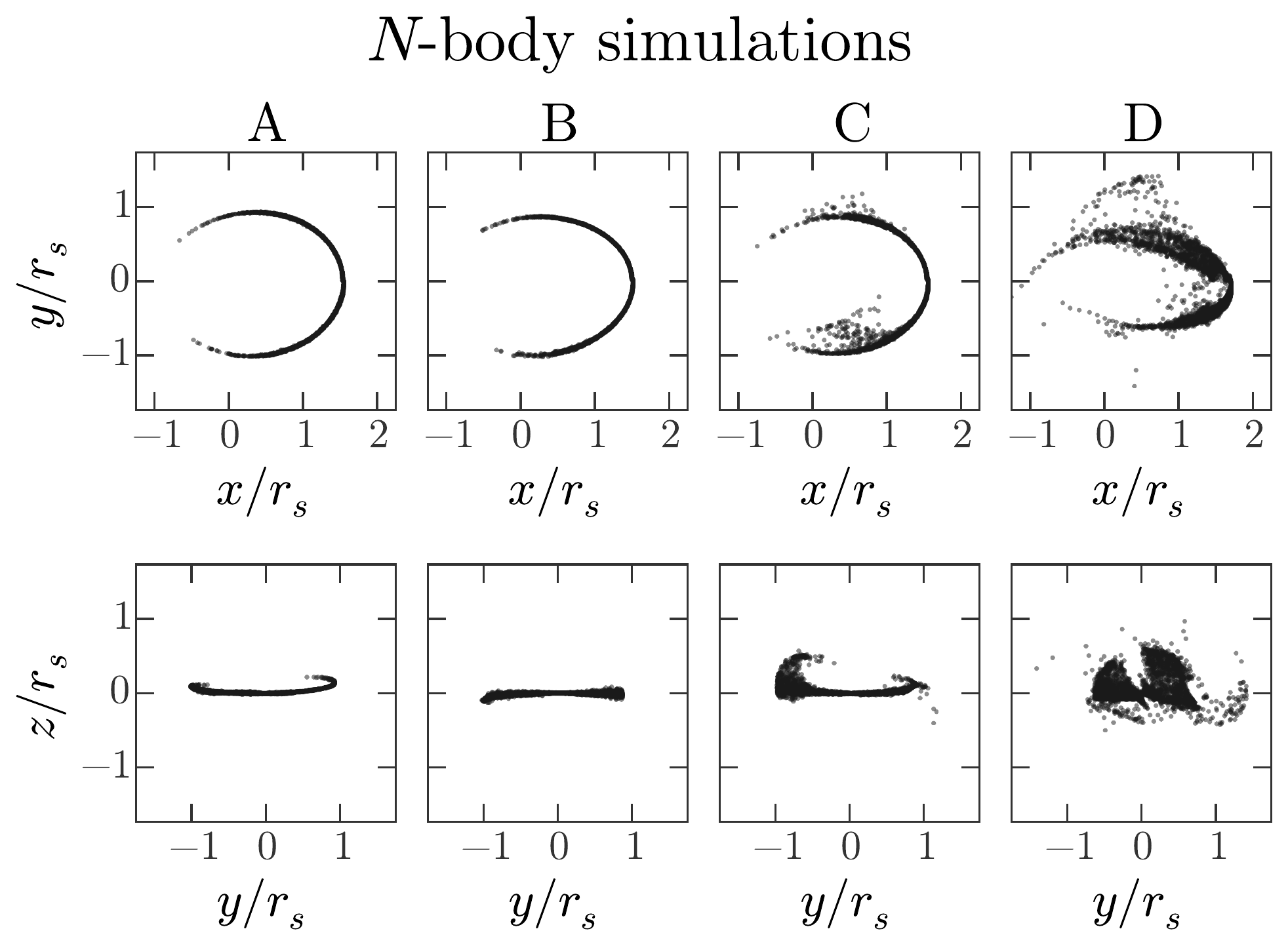}
\caption{ We use the Self-Consistent Field (SCF) basis function expansion code \citep{hernquist92} to run N-body simulations of globular-cluster-mass progenitor systems on the three orbits of Figure~\ref{fig:ensembles}. The progenitor in each simulation is initialized as a $10^4$ particle Plummer sphere and the background triaxial NFW potential is turned on slowly over 250 Myr to reduce artificial gravitational shocking. We start the progenitor systems at apocenter and integrate for $\approx$64 orbital periods so that each simulation finishes again at subsequent apocenter. The mass of the progenitor is set to $10^4~\msun$, and the length-scale of the Plummer sphere is set to 5~pc to get $\approx$50--75\% mass loss over the integration time. Panels show particle positions from the final snapshots of the simulations---for visualization the positions have been rotated so that the angular momentum of the progenitor at the final snapshot are aligned with the $z$-axis. These simulations confirm that the test-particle orbit ensembles (Figure~\ref{fig:ensembles}) do (qualitatively) capture the nature of the early-stripped tidal debris. }
\label{fig:nbodysims}
\end{center}
\end{figure}

We verify that these orbit ensembles capture the nature of more realistic stream formation by running N-body simulations of globular-cluster-mass progenitor systems on these same four orbits. We use the Self-Consistent Field (SCF) basis function expansion code \citep{hernquist92} to run the simulations, which we set up to run from apocenter-to-apocenter (rather than pericenter-to-apocenter as in the ensembles), but finish with the progenitor in the same location as the parent orbit in the ensemble evolution described above. Figure~\ref{fig:nbodysims} shows the final particle distributions rotated so that the angular momentum of the progenitor orbit is aligned with the $z$-axis. From a comparison of Figures~\ref{fig:ensembles} and \ref{fig:nbodysims}, it is clear that the morphology of the ensembles is visually similar to the `oldest' (first stripped) debris in the N-body simulations.

\begin{table*}[ht]
\begin{center}
	\begin{tabular}{c | c c c c c c | c c}
		{\bf ID} & $\bs{x/r_s}$ & $\bs{y/r_s}$ & $\bs{z/r_s}$ & $\bs{v_x/v_c}$ & $\bs{v_y/v_c}$ & $\bs{v_z/v_c}$ & $\bs{t_\lambda/\periods}$ & $\bs{t_\Omega/\periods}$ \\\toprule
A & 0.85 & 0 & 1.303 & 0 & 0.721 & 0 & $>700$ & $>10^7$ \\
\midrule
B & 0.85 & 0 & 1.152 & 0 & 0.849 & 0 & $>700$ & $>10^7$\\
\midrule
C & 0.85 & 0 & 1.27 & 0 & 0.752 & 0 & $>700$ & $\approx 3\times10^5$\\
\midrule
D & 0.85 & 0 & 1.434 & 0 & 0.597 & 0 & 8.14 & $\approx 2.5\times10^4$\\
		\bottomrule
		\end{tabular}
	\caption{Orbital initial conditions from the $xz$ grid for orbits with a range of chaotic timescales---A is a regular, near-resonant orbit, B a regular, non-resonant orbit, C a weakly chaotic orbit, and D a strongly chaotic orbit. Positions ($x$, $y$, $z$) are given in units of scale radii, $r_s$, and velocities ($v_x$, $v_y$, $v_z$) in units of scale velocity, $v_c$. Chaotic timescales are expressed ($t_\lambda$, $t_\Omega$) in number of orbital periods. Recall that the frequency diffusion time, $t_\Omega$, is the time over which we expect order-unity changes in the fundamental frequencies, hence why the timescales appear quite long. \label{tbl:orbit-info}}
\end{center}
\end{table*}

% Figure ??

Visual inspection of Figures~\ref{fig:ensembles} and \ref{fig:nbodysims} suggests that chaotic mixing of small orbit ensembles affects the configuration-space evolution of an ensemble over short times, even when the predicted chaotic timescales (from the Lyapunov and frequency diffusion times) are long: The mean, single-orbit chaos indicators are not well-suited for determining the importance of chaotic diffusion on tidal stream evolution. As a quantitative measure of this enhanced density evolution, we compute the evolution of the mean configuration-space density for orbit ensembles evolved around the four parent orbits (A,B,C,D) described above. At each time step, we use kernel density estimation (KDE)\footnote{We use an implementation from the Python package \texttt{scikit-learn} \citep{scikitlearn}.} with the ensemble of particle positions to estimate the configuration-space density field. We use an Epanechnikov kernel with an adaptive bandwidth: at each evaluation of the density, we use 10-fold cross-validation to find the optimal kernel bandwidth. We evaluate the density at the positions of each particle, $\rho_i$, and compare to the mean initial density, $\mean{\rho_0}$.

\begin{figure}[h]%[p]
\begin{center}
\includegraphics[width=0.8\textwidth]{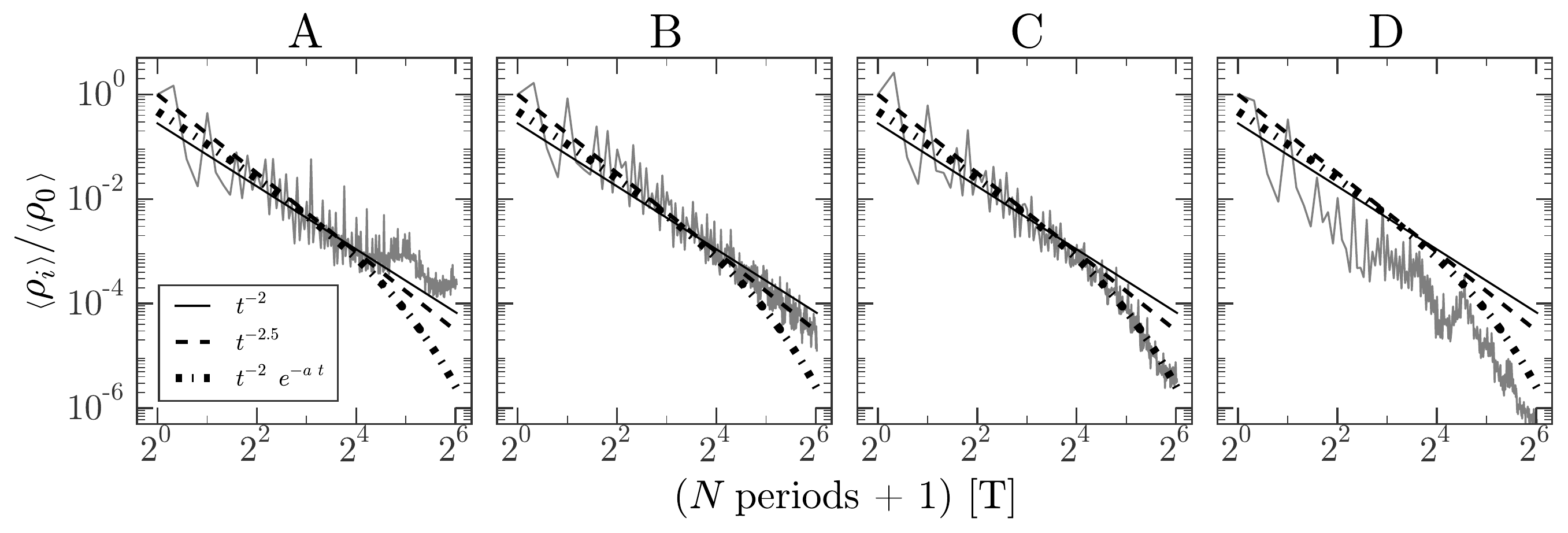}
\caption{Evolution of the mean density (normalized by the initial density) of orbit ensembles around the four orbits (Figure~\ref{fig:orbits}) over 64 orbital periods. Over-plotted are various power-laws that qualitatively show the decay of the mean density for the ensembles: the mean density of the near-resonant orbit (A) follows nearly $t^{-2}$; the non-resonant orbit (B) is slowly transitioning from $t^{-2}$ to $t^{-3}$; the weakly chaotic orbit (C) initially follows $t^{-2}$ but then exponential decay takes over; the strongly chaotic orbit (D) is roughly exponential with fairly erratic density evolution. }
\label{fig:densities}
\end{center}
\end{figure}

Figure~\ref{fig:densities} shows the mean density of the ensemble particles computed at 256 evenly-spaced intervals from the initial ensemble distribution to the distribution after 64 orbital periods. Over-laid on each panel are qualitative power-laws (i.e. not fit to the density evolution) that demonstrate the expected trends: After initial $t^{-1}$ decay, the mean density of the near-resonant orbit (A) should follow $t^{-2}$, whereas the non-resonant orbit (B) will transition from $t^{-2}$ to $t^{-3}$ after long times (and thus currently display an intermediate power-law index). Given the extremely long chaotic timescale for the weakly chaotic orbit (C), we would expect the mean density to follow a simple power-law over long times, however at $\approx$16 orbital periods the density clearly diverges and begins to follow a decaying exponential. The mean density along the strongly chaotic orbit (D) evolves stochastically but generally follows a decaying exponential.

\begin{figure}[h]%[p]
\begin{center}
\includegraphics[width=0.8\textwidth, trim={0 0.2cm 0 0}]{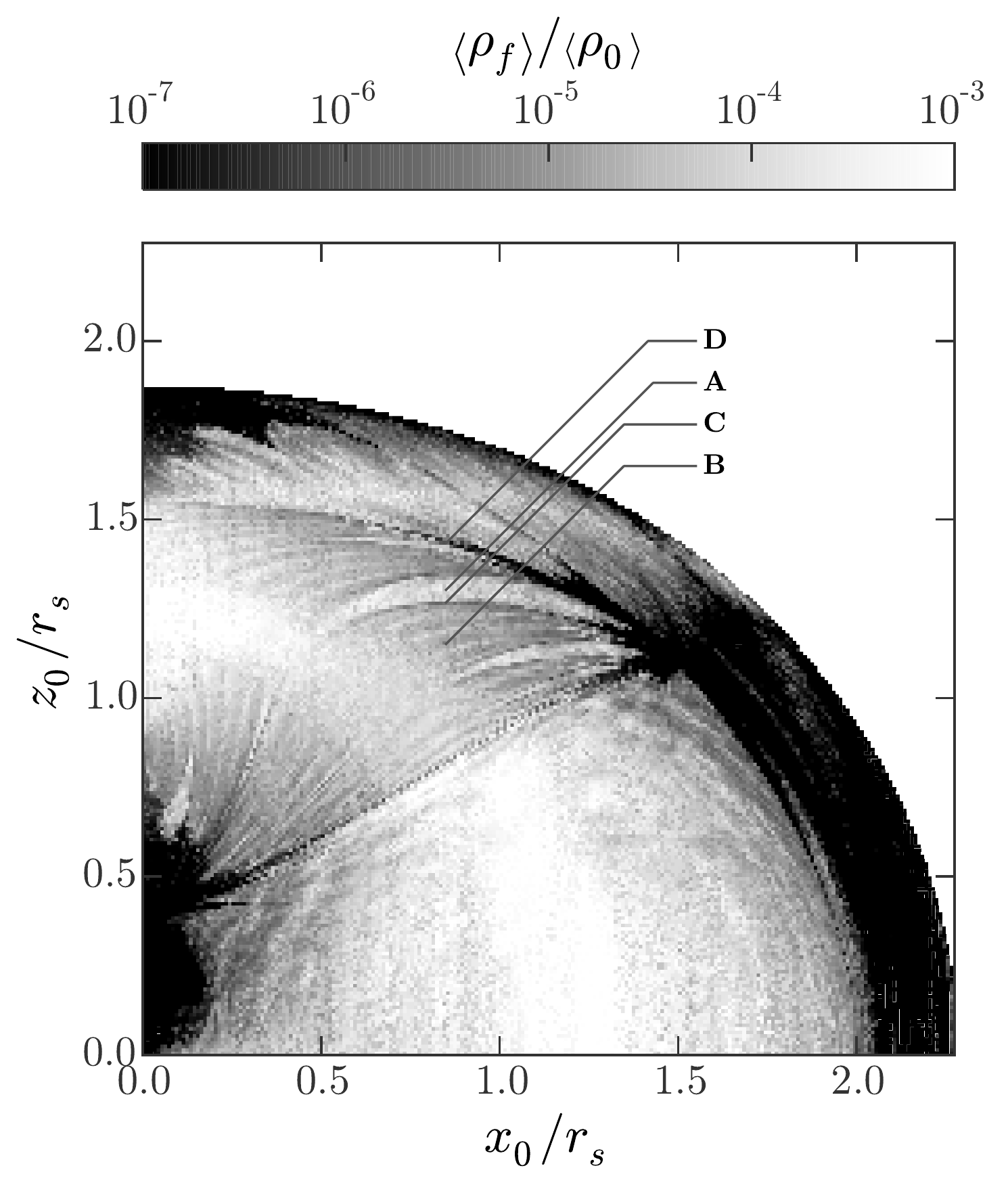}
\caption{The same grid of orbits as shown in Figure~\ref{fig:apoper}, but now the greyscale intensity is set by the mean density of a of small ensembles of orbits after integration for 64 orbital periods. The orbit ensembles are generated around each parent orbit in the grid (Section~\ref{sec:results2}) and the mean density is a kernel density estimation (KDE) with an adaptive Epanechnikov kernel. Given the extremely long chaotic timescales of the weakly chaotic orbits in the upper half of the grid, it is surprising that any of the high-order resonant structure is visible in this map. Four orbits are pointed out with arrows---from top to bottom, these are the strongly chaotic (D), near-resonant (A), weakly chaotic (C), and non-resonant (B) orbits of Table~\ref{tbl:orbit-info} (see Section~\ref{sec:results2}).}
\label{fig:ensemblemap-meandensity}
\end{center}
\end{figure}

Motivated by the noticeable discrepancy between the final mean density between the two regular (A,B) and the weakly chaotic (C) orbit ensembles---even though the chaotic timescale of the weakly chaotic ensemble parent orbit is several times the integration period used above---we compute the final mean density for orbit ensembles generated around each orbit in the grid described in Section~\ref{sec:results1}. For all ensembles, we integrate the orbits for 64 (parent) orbital periods and compute the initial and final values of the mean, configuration-space density. Figure~\ref{fig:ensemblemap-meandensity} again shows the grid of initial conditions (Section~\ref{sec:results1}, but now the greyscale indicates the ratio of the final mean density to the initial density. Interestingly, much of the structure that is visible in the upper-half of the frequency diffusion time map (Figure~\ref{fig:freqdiff}) is again visible in this map of the density evolution of orbit ensembles. Many features appear as lighter, curved features in the upper portion of this grid where the density remains systematically higher---these are the stable resonances of the potential. Darker regions have systematically lower densities and correspond to regions of resonance overlap where weakly chaotic orbits are found.

% Figure ??
\clearpage

However, it is surprising that these features are present: The chaotic timescales predicted from both the Lyapunov and frequency diffusion times were typically 100 to 1000s of orbital periods for many of the unstable features around resonances. The timescales in the lower portion of the grid (where the orbits are primarily stable, short-axis tube orbits) are simply too long to detect density differences over 64 orbital periods even from weak chaos in this particular choice of potential.

It is worth noting that the frequency diffusion time map predicts that the resonances are surrounded by thick stochastic layers, while in the density map the resonance layers appear to be stable (and thus lead to higher mean densities). This comes from a failure of frequency determination: Near resonances it can take extremely long integration periods to ensure accurate recovery of the frequencies due to aliasing \citep{laskar03}. However, at the edges of the resonance layers---where we expect to find stochastic layers and do see weakly chaotic motion---the frequency diffusion time is a robust indicator of chaos.

We conclude from these experiments that the degree of chaos is an indicator that ensembles of orbits may mix faster than predicted from regular phase-mixing, however the Lyapunov time and frequency diffusion time are not good predictors for the timescale over which this mixing will occur. To understand this discrepancy, we next explore why this occurs (Section~\ref{sec:results3}, in the context of Section~\ref{sec:chaotic-mixing}).

% Figure ??
\clearpage

\subsection{Part III: Short-time frequency evolution}\label{sec:results3}

Why does chaos manifest itself after just 64 orbital periods around orbits with predicted chaotic timescales larger than thousands of orbital periods? Both the Lyapunov exponent and the frequency diffusion rate measure mean, long-term rates of chaotic diffusion. If a weakly chaotic orbit is confined (by other nearby, non-overlapping resonances or stable regions), the mean drift of an orbit in frequency space may be small if computed over timescales long compared to the orbital time but short compared to the Arnold diffusion time, however small-scale variation of the frequencies may occur over much shorter timescales.

\begin{figure}[h]%[p]
\begin{center}
\includegraphics[width=\textwidth]{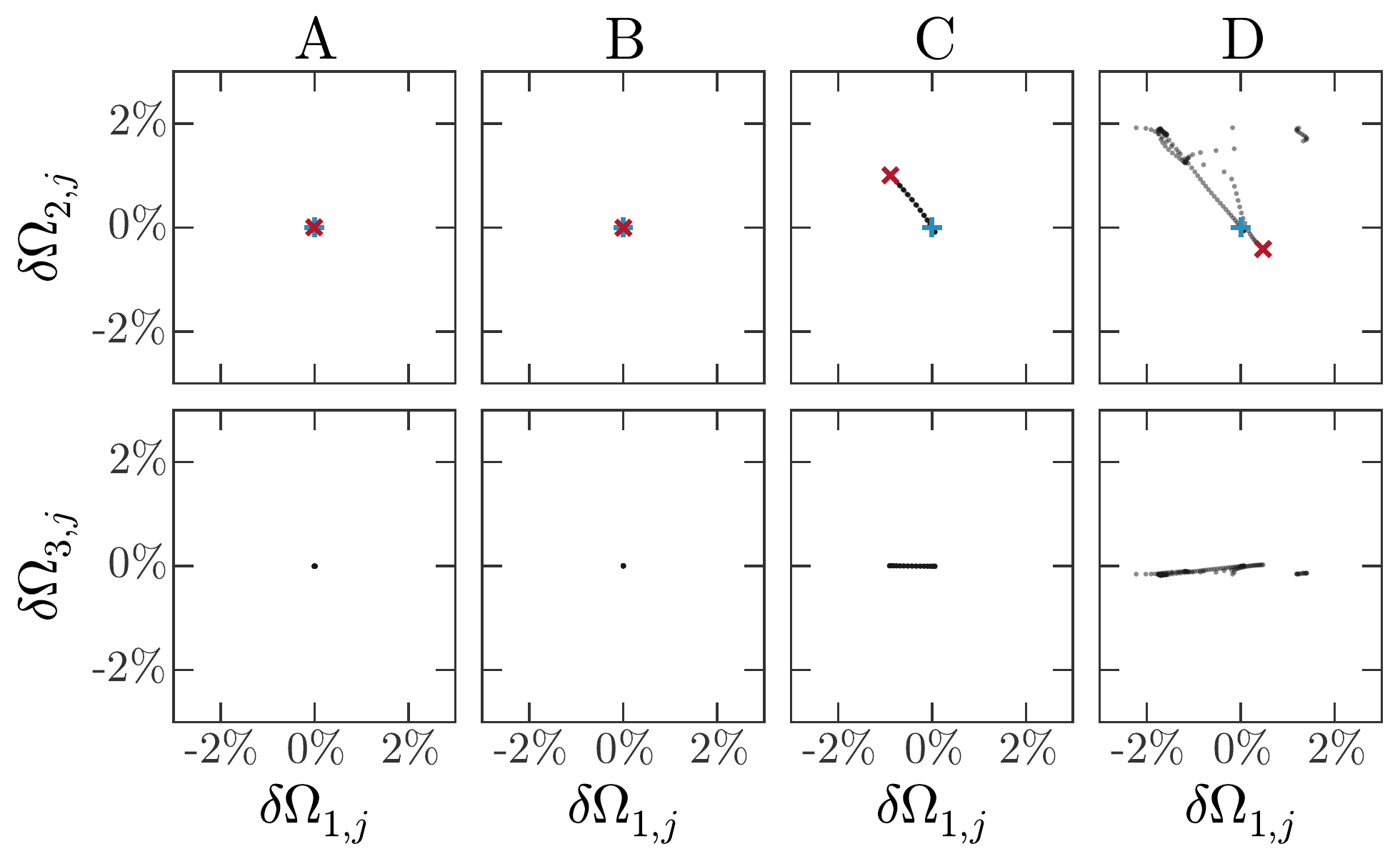}
\caption{Evolution of the fundamental frequencies computed over a total of 256 orbital periods for the four orbits, A, B, C, D. Plotted are the per cent deviations of the frequencies computed in window relative to the value in the initial window---that is, if $j$ is the index of a given time window, $\delta \Omega_{1,j} = (\Omega_{1,j} - \Omega_{1,0})/\Omega_{1,0}$. The initial value is shown as a blue plus sign and the final value is shown as a red x. The frequencies are computed with a window width of $\approx$128 orbital periods, and the window is shifted by one orbital period between each computation (each grey point represents the fundamental frequencies computed in a single window). For the regular orbits (A,B), the fractional variation is around $10^{-6}$ and thus all points overlap on this scale. The weakly chaotic orbit (C) displays frequency variations comparable to the spread in frequencies in globular-cluster-like tidal debris. The frequency variations of the strongly chaotic orbit (D) have a larger characteristic spread.}
\label{fig:four-orbits-freqs}
\end{center}
\end{figure}

As a demonstration of the small-scale frequency modulation, we consider again the four orbits of Section~\ref{sec:results2} (initial conditions are listed in Table~\ref{tbl:orbit-info}). To resolve the short-time behavior of the frequency diffusion (corresponding to motion across or around resonance layers) we compute the frequencies in a series of rolling windows along each orbit. We use a window with a width equal to 64 orbital periods and shift the window by an orbital period between each calculation of the most significant frequencies. Figure~\ref{fig:four-orbits-freqs} shows plots of the frequencies of each orbit computed in each window of time---plotted in projection are the percent deviations of the frequency from the initial value. The difference between the first window (blue +) and the last window (red x) is 128 orbital periods. For the two regular orbits, (A) and (B), the frequencies vary only from numerical issues when recovering the frequencies ($\ll$1\%) and thus the initial and final value markers overlap. The frequencies of the weakly chaotic orbit (C) evolve quickly but are bounded to a small volume with fractional size $\approx$1\% (presumably by nearby resonant surfaces); this is larger than the frequency spread in globular cluster tidal debris (0.1--0.5\%). The frequencies of the strongly chaotic orbit (D) also evolve quickly but fill a larger volume.

We see now where the discrepancy between chaotic timescales and the observable effects of chaos arise in tidal debris: even if the large-scale diffusion of frequencies is slow, an orbit may explore a region of frequency space over much shorter times. The frequency diffusion time (Equation~\ref{eq:fdtime}) is an estimate of the time over which the mean value of the frequencies evolves, however the \emph{variance} of the frequencies over short times is dynamically relevant for small ensembles. This small-scale variability is insignificant for the evolution of global structure in, for example, an elliptical galaxy or in erasing substructure in the Solar neighborhood \citep[e.g.,][]{maffione15}, but is signifiant for the morphological evolution of tidal debris with small spreads in frequencies.

We therefore expect a small ensemble of orbits in frequency space to expand over short times around even weakly chaotic parent orbits and the debris should therefore appear dynamically hotter in real-space. The effect is especially significant if the chaotic evolution of the frequencies occurs orthogonal to the largest eigenvector of the distribution of frequencies or the local Hessian of the Hamiltonian. We illustrate this phenomenon by computing the frequencies for each orbit in small ensembles of orbits around each of the four orbits (A,B,C,D).

\begin{figure}[h]%[p]
\begin{center}
\includegraphics[width=\textwidth]{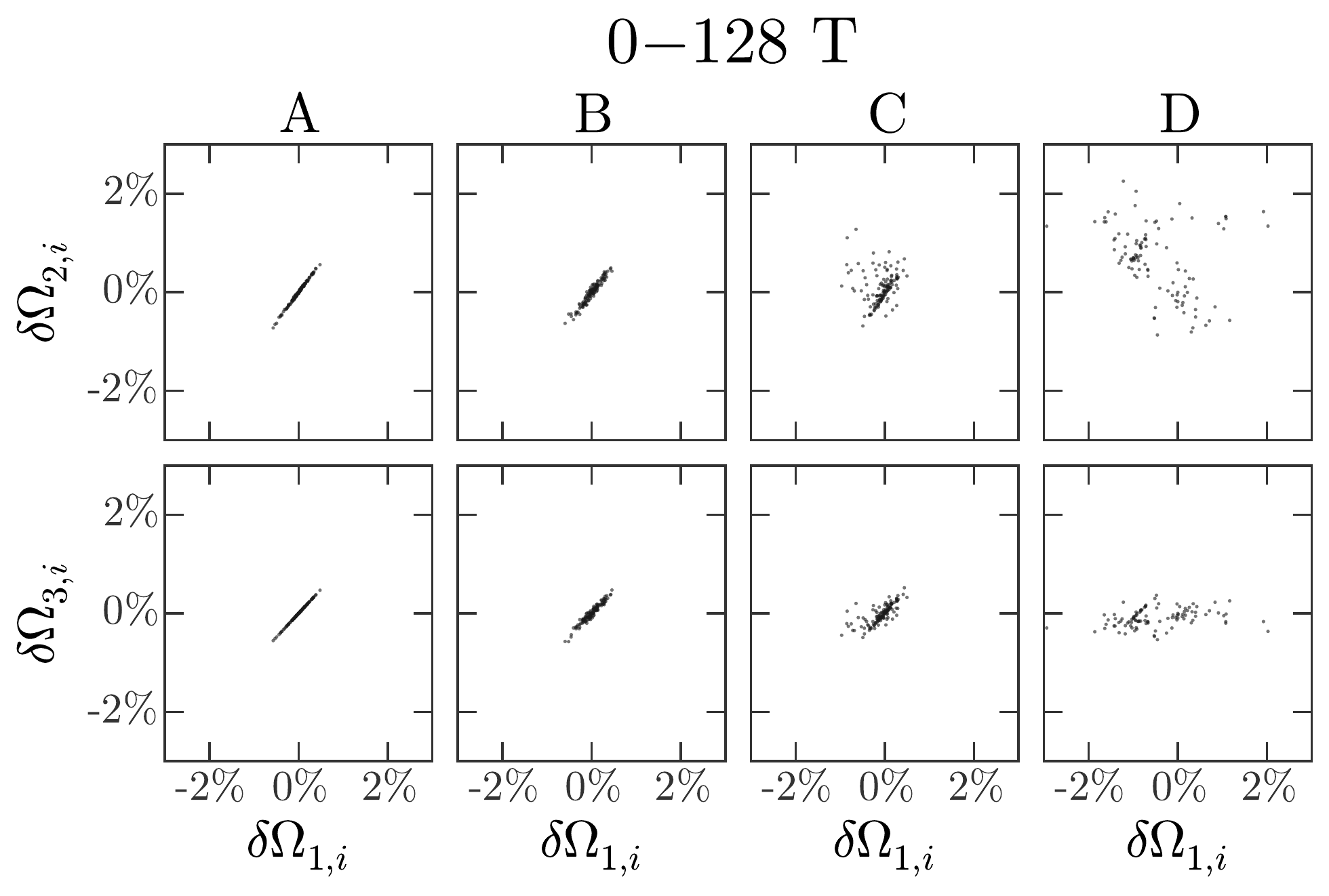}
\caption{Distributions of the fundamental frequencies for all orbits in 128-orbit ensembles generated around the four orbits, A, B, C, D. Plotted are the per cent deviations of the frequencies of each ensemble orbit relative to the frequencies of the parent orbit---that is, if $i$ is the index of a given orbit and $i=0$ is the parent orbit, $\delta \Omega_{1,i} = (\Omega_{1,i} - \Omega_{1,0})/\Omega_{1,0}$. All orbits are integrated for 128 orbital periods to compute the frequencies. The near-resonant ensemble frequency distribution (A) is nearly 1D, and thus the ensemble appears one-dimensional in configuration-space (Figure~\ref{fig:ensembles}). The two largest eigenvalues of the non-resonant ensemble frequency distribution (B) are closer, and thus the ensemble spreads quicker, leading to a two-dimensional spread of debris in configuration-space. Around both the weakly chaotic orbit (C) and strongly chaotic orbit (D), chaotic diffusion increases the spread of the debris significantly, leading to faster 3D spreading in configuration-space.}
\label{fig:ensemble-freqs0}
\end{center}
\end{figure}

\begin{figure}[h]%[p]
\begin{center}
\includegraphics[width=\textwidth]{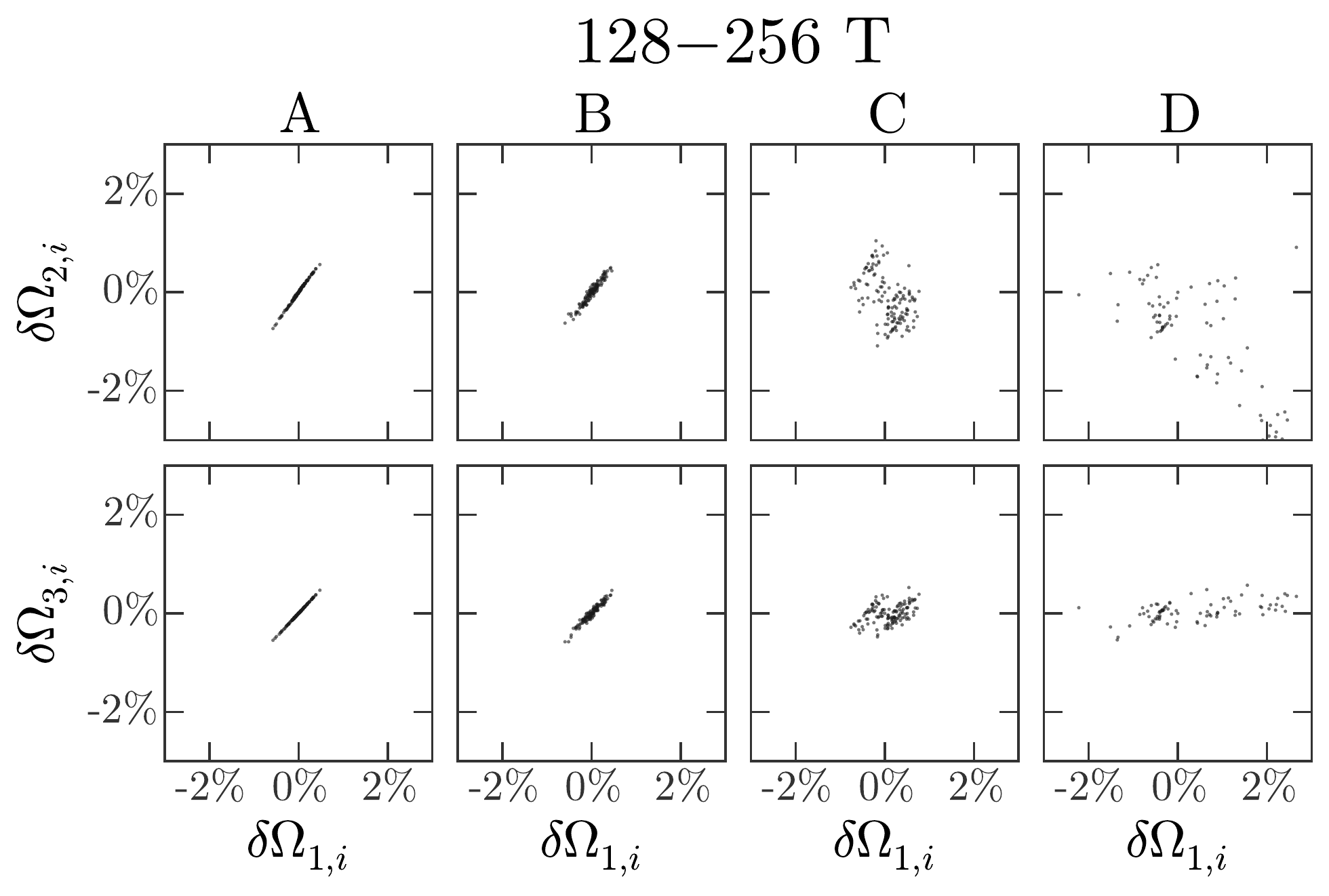}
\caption{Same as Figure~\ref{fig:ensemble-freqs0} but after integrating the same orbits for another 128 orbital periods. The ensemble frequency distributions around the regular orbits (A,B) appear nearly identical, whereas both the weakly and strongly chaotic ensemble frequency distributions spread significantly.}
\label{fig:ensemble-freqs1}
\end{center}
\end{figure}

We generate orbit ensembles with 128 orbits around the four orbits and integrate for two consecutive windows of 128 orbital periods. Figures~\ref{fig:ensemble-freqs0} and \ref{fig:ensemble-freqs1} show the distribution of frequencies for all orbits in the ensembles for the four orbits in each of the two consecutive windows. Around the near-resonant orbit (A), the frequency distribution is nearly one-dimensional, which explains the thin, one-dimensional morphology of the ensemble in the left column of Figure~\ref{fig:ensembles}. The non-resonant orbit ensemble (B) is `thicker'---the ratio of the two largest eigenvalues of this distribution is larger---and therefore in real space, the ensemble begins to spread over two dimensions, as is also evident in the final ensemble debris morphology in the middle-left panel of Figure~\ref{fig:ensembles}. Non-resonant orbits in axisymmetric or triaxial potentials will generically have frequency distributions that have more comparable spreads in each direction and the configuration-space density of typical tidal debris will therefore decrease faster in such potentials \citep{helmi99}. The weakly chaotic orbit ensemble (C) is clearly more diffuse than the two regular ensembles: As is evident in Figure~\ref{fig:four-orbits-freqs}, the small-scale chaotic evolution of the frequencies---though bounded---increases the variance of the frequency distribution. By comparing Figure~\ref{fig:ensemble-freqs0} and \ref{fig:ensemble-freqs1}, it appears that the frequency distribution gets more diffuse until it uniformly fills the allowed volume. The strongly chaotic orbit ensemble (D) already has a large spread in Figure~\ref{fig:ensemble-freqs0}, and in Figure~\ref{fig:ensemble-freqs1} many of the frequencies have fractional frequency differences relative to the parent orbit $\approx$5--10\% (off of the plot).

% Check frequency analysis precision
\clearpage

% Figure - Ensemble frequencies
\clearpage

% Figure - Ensemble frequencies
\clearpage

\section{Discussion: limitations and future work}\label{sec:discussion}

% Something about this: The timescales in this paper are not meant to be interpreted physically, as the potential is a very crude approximation to the total potential of a Milky-Way-like galaxy. The key point is that even when the mean chaos indicators predict a long chaotic timescale, streams may display enhanced density evolution over much shorter times. Tidal debris is sensitive to the small-scale, short-time evolution of orbital properties (e.g., the fundamental frequencies) induced by weak chaos.

We have shown that the Lyapunov and frequency diffusion times are indicators of chaos and that the frequency diffusion time resolves the detailed resonant structure of gravitational potentials, but the timescales predicted do not capture the importance of chaos for the density evolution of ensembles of orbits meant to mimic tidal debris. We have shown that small-scale but fast chaotic diffusion of frequencies can explain this enhanced mixing. In the sub-sections below, we discuss a few important limitations that remain for exploration in future work.

\subsection{The progenitor mass scale}

We have only considered low-mass progenitor systems such as globular clusters because the intrinsic spreads in fundamental frequencies are small (0.1--0.5\%). Small changes to the frequencies of the orbits of tidal debris stripped from these progenitors due to chaotic processes will therefore cause observable changes to the real-space morphology of the debris. For more massive progenitor systems, the typical size and velocity dispersion of the debris will be larger and thus the debris morphology will be less sensitive to small changes in orbital properties. The mass-scale of the debris that will display enhanced density evolution depends on the magnitude of weak chaos, which depends on the orbit structure of a given potential. In potentials with more significant chaos, debris disrupted from more massive progenitor systems may also display `stream-fanning.' However, since the scale of the debris is larger it is more likely that some orbits will exist in chaotic regions.

\subsection{Potential choice}

The potential considered in this work is `unrealistic' for the total potential of a Milky-Way-like galaxy in that it is static, smooth, and does not contain baryonic components. Simulations of forming galactic disks in cosmological dark-matter haloes have shown that baryonic feedback and relaxation can significantly change the inner distribution of dark matter and either make the potential more spherical or oblate \citep{dubinski94,kazantzidis04, bryan13, butsky15}. However, the significance of baryonic relaxation or of time-dependence, triaxiality, and substructure on shaping the matter distribution within the Milky Way is largely unknown. Here we briefly summarize future directions for potential models:
\begin{itemize}
	\item \emph{Baryonic components}: \cite[][D08]{debattista08} and \cite[][V10]{valluri10} studied the orbit evolution induced by growing a baryonic disk in dark-matter haloes with various shapes and orientations (e.g., prolate, triaxial). In general, the authors find that the growth of a disk slowly deforms the orbits of mass tracers (e.g., dark matter particles) and preferentially populates tube or other `round' orbits (i.e. even the chaotic and box orbits fill a fairly spherical or oblate volume). Consequently, the inner shape of the potential becomes more oblate or spherical. If the inner haloes of galaxies are indeed close to spherical or oblate, the majority of orbits will be regular and chaos will be less important, however this is far from conclusive. We have found from simple experiments in superposing potential components that transition regions between potential shapes can significantly enhance the amount of chaos. Though superposing potentials is not realistic, it at least suggests that a more careful exploration of potential configurations is required to understand how complex, radius-dependent potential forms affect the amount and significance of chaos in galactic haloes.

\item \emph{Time dependence}: Galaxies are certainly not static systems. To first order, galaxies grow in mass---for example, the spherically averaged mass profile of dark-matter haloes evolves fairly predictably in cosmological simulations \citep{wechsler02, buist14} after some initial period of stochastic mass growth. The Milky Way has probably had a fairly calm accretion history over the last 6 Gyr and therefore the mass growth may be similar to that seen in simulations. This steady growth most likely does not alter the global structure or shape of the potential. However, we also know from simulations that figure rotation, baryonic feedback, and the accretion and phase-mixing of subhaloes do perturb the global state of simulated haloes. \cite{deibel11} showed that by adopting pattern speeds comparable to those found in cosmological simulations, figure rotation generally acts to destabilize orbits (rather than stabilize chaotic orbits in the equivalent static potential) and the resulting orbit structure is that most regular orbits are associated with resonances. The presence of and response to the Large and Small Magellanic Clouds may also introduce significant (time-dependent) perturbations to the global potential of the Milky Way \citep[e.g.,][]{besla10, gomez15}. In future work we will explore the effect of these time-dependent processes on the chaotic dispersal of tidal streams using live potentials from cosmological N-body simulations.

\item \emph{Substructure}: Cosmological simulations predict that dark-matter haloes are filled with substructure in the form of dark matter subhaloes. If they exist, these subhaloes may account for up to $\approx$1-10\% of the mass of the dark matter \citep[e.g.,][]{diemand07} and therefore may contribute significantly to and orbit in the large-scale potential of any galaxy. Gravitational scattering due to subhalo interactions has been studied, however in the haloes of galaxies where dynamical times are long, the scattering cross-section for \emph{strong} encounters is small. Instead, the collective effect of the subhaloes may instead act as a noise term in the Hamiltonian of any halo orbit. This subhalo-induced heating---which depends on the mass spectrum and distribution of subhaloes---may also act to simply increase the magnitude of chaos along orbits, and destabilize sufficiently non-resonant orbits \citep[see, e.g.,][]{kandrup00, siegalgaskins08}.

\end{itemize}

\subsection{Stream modeling}

Tidal stream modeling is one of the most promising ways to constrain the 3D mass distribution around the Milky Way at distances of 10s to 100s of kpc. Methods that use tidal streams to infer properties of the Galactic potential typically operate by constructing models of the debris distribution using either the present-day phase-space density \citep[e.g.,][]{varghese11, kuepper12, kuepper15, gibbons14}, time-of-disruption phase-space density \citep{apw13, apw14}, or the density in angle-action coordinates \citep{sanders14, bovy14}. All of these methods may fail or produce uninterpretable results if modeling globular-cluster streams on mildly chaotic orbits. For each of these methods, it is important to understand the failures and biases introduced by ignoring chaotic orbit evolution.

\subsection{Ophiuchus stream}

The Ophiuchus stream \citep{bernard14, sesar15a} appears to be a thin, short tidal stream (deprojected length $\approx$1.5--2 kpc) near the Galactic bulge with no apparent progenitor. At Galactocentric $R \approx 1$ kpc, $z \approx 5$ kpc, the orbits of the stream stars likely pass through the MW disk, feel the triaxiality of the Galactic bar \citep[e.g.,][]{wegg13, wegg15}, and have short orbital periods (relative to streams in the halo). It is possible that the observed debris is the last remnants of the recently disrupted progenitor system \citep{sesar15a}, however if a significant number of stars were stripped on previous pericentric passages, this older debris may be `fanned' and still near the observed portion of the stream. The fanned debris would have significantly lower surface brightness and thus would be more difficult to detect. The detection of this low-surface-brightness component would open up the possibility that enhanced density evolution due to chaos is a dynamically relevant process for thin streams in real galaxies, though \cite{carlberg15} has shown that it is also possible to get short, high density segments of streams from debris formed on eccentric orbits.

\section{Summary and Conclusions}\label{sec:conclusions}

We have considered here a simple triaxial gravitational potential chosen to mimic the median properties of dark-matter haloes formed in dark-matter-only simulations (or the large-scale properties of haloes formed in simulations with baryonic effects). We have numerically computed the magnitude of chaos for a large grid of iso-energy orbits in this potential using two independent methods that have been used extensively to classify and characterize chaotic orbits: 1) the Lyapunov exponent and 2) the frequency diffusion rate. From each of these indicators, we compute a timescale over which chaos is likely to be important and find that the majority of orbits have chaotic timescales greater than 100s of orbital periods, however with the frequency diffusion rate we are still able to resolve weak chaos. We then study the density evolution of small ensembles of orbits generated around each orbit in the grid used for the previous experiment and find that along some orbits classified as weakly chaotic---with chaotic timescales of 100s of orbital periods---the orbit ensembles display enhanced density evolution and reach a lower overall density faster than orbit ensembles around nearby regular orbits (which mix due to phase-mixing alone). We explain this discrepancy between the predicted chaotic timescale and the observed effects of chaos on tidal debris by considering the nature of chaotic diffusion: the classical chaos indicators are most sensitive to the slow, Arnold diffusion process that can cause large changes to orbital properties, but small-scale frequency evolution occurs over much shorter times as chaotic orbits stochastically diffuse across the stochastic layers that surround many resonances. The fundamental frequencies of a weakly chaotic orbit therefore vary over a small region (bounded by nearby stable resonances), and when this amplitude is comparable to or larger than the typical spread in frequencies of tidal debris from the progenitor, the phase-space density of the debris will evolve faster to a state of lower density relative to nearby regular orbits.

Our main results and conclusions are summarized as follows:
\begin{enumerate}
	\item ``stream-fanning'' of tidal streams on weakly chaotic orbits \citep[as seen in simulations by][]{pearson15} occurs due to small-scale chaotic evolution of the fundamental frequencies of the debris star orbits;
	\item the Lyapunov time and frequency diffusion time are powerful indicators of chaos, but do not capture the importance of small-scale chaotic diffusion for the density evolution of small ensembles of orbits (tidal debris);
	\item tidal debris becomes diffuse and thus harder to observe on weakly chaotic orbits when the small-scale chaotic diffusion of the fundamental frequencies has a scale comparable to the internal spread in frequencies of the debris;
	\item the details of the enhanced mixing along weakly chaotic orbits depends on the resonant structure of the potential;
	\item the covariance of the fundamental frequencies of an orbit over a given time window may be a better predictor of the importance of small-scale chaotic frequency diffusion on the resulting morphology of tidal debris.
\end{enumerate}

Our results provide a clear explanation of how and why the morphology of tidal streams alone can be used to constrain the potential of the host galaxy. The longest thin streams are most valuable for this effort because they have clearly evolved for a long time, but the debris remains compact. For shorter thin streams it will be hard to decouple the unknown evolution time from enhanced density evolution from chaos. The mere existence of thin tidal streams in the halo of the Milky Way either (1) provides useful information about the potential on these scales by, e.g., implying a large degree of regularity, or (2) indicates that the thin, long streams (e.g., GD-1) are on regular orbits. These are not mutually exclusive---in fact, if the streams are on regular orbits, this would be a powerful way to check or rule out possible potential models by requiring that the progenitor orbits remain regular.

\acknowledgements
The authors wish to thank Robyn Sanderson, Dan D'Orazio, David Merritt, and the \emph{Stream Team} for useful comments and discussion. \chchchanges{The authors additionally thank the referee, Walter Dehnen, for insightful comments that greatly improved this article.}

APW is supported by a National Science Foundation Graduate Research Fellowship under Grant No.\ DGE 11-44155. KVJ and APW were partially supported by the National Science Foundation under Grant No. AST-1312196 and NASA grant NNX15AK78G. MV is supported in part by the University of Michigan's Elizabeth Crosby Fund and the Office of the VP for Research and NASA ATP grant NNX15AK79G. AHWK would like to acknowledge support from NASA through Hubble Fellowship grant HST-HF-51323.01-A awarded by the Space Telescope Science Institute, which is operated by the Association of Universities for Research in Astronomy, Inc., for NASA, under contract NAS 5-26555.

This research made use of \project{Astropy}, a community-developed core \project{Python} package for Astronomy \citep{astropy13}. This work additionally relied on Columbia University's \emph{Hotfoot} and \emph{Yeti} compute clusters, and we acknowledge the Columbia HPC support staff for assistance. This work used the Extreme Science and Engineering Discovery Environment (XSEDE), which is supported by National Science Foundation grant number ACI-1053575 \citep{xsede}.

\bibliographystyle{apj}
\bibliography{refs}

\begin{thebibliography}{}
\expandafter\ifx\csname natexlab\endcsname\relax\def\natexlab#1{#1}\fi

\bibitem[{{Arnold}(1978)}]{arnold78}
{Arnold}, V.~I. 1978, {Mathematical methods of classical mechanics}

\bibitem[{{Astropy Collaboration} {et~al.}(2013){Astropy Collaboration},
  {Robitaille}, {Tollerud}, {Greenfield}, {Droettboom}, {Bray}, {Aldcroft},
  {Davis}, {Ginsburg}, {Price-Whelan}, {Kerzendorf}, {Conley}, {Crighton},
  {Barbary}, {Muna}, {Ferguson}, {Grollier}, {Parikh}, {Nair}, {Unther},
  {Deil}, {Woillez}, {Conseil}, {Kramer}, {Turner}, {Singer}, {Fox}, {Weaver},
  {Zabalza}, {Edwards}, {Azalee Bostroem}, {Burke}, {Casey}, {Crawford},
  {Dencheva}, {Ely}, {Jenness}, {Labrie}, {Lim}, {Pierfederici}, {Pontzen},
  {Ptak}, {Refsdal}, {Servillat}, \& {Streicher}}]{astropy13}
{Astropy Collaboration}, {Robitaille}, T.~P., {Tollerud}, E.~J., {et~al.} 2013,
  \aap, 558, A33

\bibitem[{{Bailin} {et~al.}(2005){Bailin}, {Kawata}, {Gibson}, {Steinmetz},
  {Navarro}, {Brook}, {Gill}, {Ibata}, {Knebe}, {Lewis}, \&
  {Okamoto}}]{bailin05}
{Bailin}, J., {Kawata}, D., {Gibson}, B.~K., {et~al.} 2005, \apjl, 627, L17

\bibitem[{{Belokurov} {et~al.}(2006){Belokurov}, {Zucker}, {Evans}, {Gilmore},
  {Vidrih}, {Bramich}, {Newberg}, {Wyse}, {Irwin}, {Fellhauer}, {Hewett},
  {Walton}, {Wilkinson}, {Cole}, {Yanny}, {Rockosi}, {Beers}, {Bell},
  {Brinkmann}, {Ivezi{\'c}}, \& {Lupton}}]{belokurov06}
{Belokurov}, V., {Zucker}, D.~B., {Evans}, N.~W., {et~al.} 2006, \apjl, 642,
  L137

\bibitem[{Benettin {et~al.}(1976)Benettin, Galgani, \& Strelcyn}]{benettin76}
Benettin, G., Galgani, L., \& Strelcyn, J.-M. 1976, Phys. Rev. A, 14, 2338

\bibitem[{{Bernard} {et~al.}(2014){Bernard}, {Ferguson}, {Schlafly}, {Abbas},
  {Bell}, {Deacon}, {Martin}, {Rix}, {Sesar}, {Slater}, {Pe{\~n}arrubia},
  {Wyse}, {Burgett}, {Chambers}, {Draper}, {Hodapp}, {Kaiser}, {Kudritzki},
  {Magnier}, {Metcalfe}, {Morgan}, {Price}, {Tonry}, {Wainscoat}, \&
  {Waters}}]{bernard14}
{Bernard}, E.~J., {Ferguson}, A.~M.~N., {Schlafly}, E.~F., {et~al.} 2014,
  \mnras, 443, L84

\bibitem[{{Besla} {et~al.}(2010){Besla}, {Kallivayalil}, {Hernquist}, {van der
  Marel}, {Cox}, \& {Kere{\v s}}}]{besla10}
{Besla}, G., {Kallivayalil}, N., {Hernquist}, L., {et~al.} 2010, \apjl, 721,
  L97

\bibitem[{{Bett} {et~al.}(2007){Bett}, {Eke}, {Frenk}, {Jenkins}, {Helly}, \&
  {Navarro}}]{bett07}
{Bett}, P., {Eke}, V., {Frenk}, C.~S., {et~al.} 2007, \mnras, 376, 215

\bibitem[{{Binney} \& {Tremaine}(2008)}]{binneytremaine}
{Binney}, J., \& {Tremaine}, S. 2008, {Galactic Dynamics: Second Edition}
  (Princeton University Press)

\bibitem[{{Bonaca} {et~al.}(2012){Bonaca}, {Geha}, \&
  {Kallivayalil}}]{bonaca12}
{Bonaca}, A., {Geha}, M., \& {Kallivayalil}, N. 2012, \apjl, 760, L6

\bibitem[{{Bovy}(2014)}]{bovy14}
{Bovy}, J. 2014, \apj, 795, 95

\bibitem[{{Bryan} {et~al.}(2013){Bryan}, {Kay}, {Duffy}, {Schaye}, {Dalla
  Vecchia}, \& {Booth}}]{bryan13}
{Bryan}, S.~E., {Kay}, S.~T., {Duffy}, A.~R., {et~al.} 2013, \mnras, 429, 3316

\bibitem[{{Buist} \& {Helmi}(2014)}]{buist14}
{Buist}, H.~J.~T., \& {Helmi}, A. 2014, \aap, 563, A110

\bibitem[{{Butsky} {et~al.}(2015){Butsky}, {Macci{\`o}}, {Dutton}, {Wang},
  {Stinson}, {Penzo}, {Kang}, {Keller}, \& {Wadsley}}]{butsky15}
{Butsky}, I., {Macci{\`o}}, A.~V., {Dutton}, A.~A., {et~al.} 2015, ArXiv
  e-prints, arXiv:1503.04814

\bibitem[{{Carlberg}(2015)}]{carlberg15}
{Carlberg}, R.~G. 2015, \apj, 808, 15

\bibitem[{{Chirikov}(1960)}]{chirikov60}
{Chirikov}, B.~V. 1960, Journal of Nuclear Energy, 1, 253

\bibitem[{{Chirikov}(1979)}]{chirikov79}
---. 1979, \physrep, 52, 263

\bibitem[{{de Zeeuw}(1985)}]{deZeeuw85}
{de Zeeuw}, T. 1985, \mnras, 216, 273

\bibitem[{{Debattista} {et~al.}(2008){Debattista}, {Moore}, {Quinn},
  {Kazantzidis}, {Maas}, {Mayer}, {Read}, \& {Stadel}}]{debattista08}
{Debattista}, V.~P., {Moore}, B., {Quinn}, T., {et~al.} 2008, \apj, 681, 1076

\bibitem[{{Deibel} {et~al.}(2011){Deibel}, {Valluri}, \& {Merritt}}]{deibel11}
{Deibel}, A.~T., {Valluri}, M., \& {Merritt}, D. 2011, \apj, 728, 128

\bibitem[{{Diemand} {et~al.}(2007){Diemand}, {Kuhlen}, \& {Madau}}]{diemand07}
{Diemand}, J., {Kuhlen}, M., \& {Madau}, P. 2007, \apj, 657, 262

\bibitem[{{Dubinski}(1994)}]{dubinski94}
{Dubinski}, J. 1994, \apj, 431, 617

\bibitem[{{Fardal} {et~al.}(2015){Fardal}, {Huang}, \& {Weinberg}}]{fardal14}
{Fardal}, M.~A., {Huang}, S., \& {Weinberg}, M.~D. 2015, \mnras, 452, 301

\bibitem[{{Gibbons} {et~al.}(2014){Gibbons}, {Belokurov}, \&
  {Evans}}]{gibbons14}
{Gibbons}, S.~L.~J., {Belokurov}, V., \& {Evans}, N.~W. 2014, \mnras, 445, 3788

\bibitem[{Goldstein(1980)}]{goldstein80}
Goldstein, H. 1980, Classical Mechanics, Addison-Wesley series in physics
  (Addison-Wesley Publishing Company)

\bibitem[{{G{\'o}mez} {et~al.}(2015){G{\'o}mez}, {Besla}, {Carpintero},
  {Villalobos}, {O'Shea}, \& {Bell}}]{gomez15}
{G{\'o}mez}, F.~A., {Besla}, G., {Carpintero}, D.~D., {et~al.} 2015, \apj, 802,
  128

\bibitem[{{Grillmair}(2006)}]{grillmair06b}
{Grillmair}, C.~J. 2006, \apjl, 645, L37

\bibitem[{{Grillmair} \& {Dionatos}(2006)}]{grillmair06a}
{Grillmair}, C.~J., \& {Dionatos}, O. 2006, \apjl, 643, L17

\bibitem[{Hairer {et~al.}(1993)Hairer, N{\o}rsett, \& Wanner}]{hairer93}
Hairer, E., N{\o}rsett, S.~P., \& Wanner, G. 1993, Springer Series in
  Computational Mathematics, Vol.~8, Solving Ordinary Differential Equations.
  {I}. Nonstiff Problems, 2nd edn. (Berlin: Springer-Verlag), xvi+528, a
  reprinting with corrections appeared in 2000.

\bibitem[{{Helmi} \& {White}(1999)}]{helmi99}
{Helmi}, A., \& {White}, S.~D.~M. 1999, \mnras, 307, 495

\bibitem[{{Hernquist} \& {Ostriker}(1992)}]{hernquist92}
{Hernquist}, L., \& {Ostriker}, J.~P. 1992, \apj, 386, 375

\bibitem[{{Hunter}(2002)}]{hunter02}
{Hunter}, C. 2002, \ssr, 102, 83

\bibitem[{{Ibata} {et~al.}(1994){Ibata}, {Gilmore}, \& {Irwin}}]{ibata94}
{Ibata}, R.~A., {Gilmore}, G., \& {Irwin}, M.~J. 1994, \nat, 370, 194

\bibitem[{{Jing} \& {Suto}(2002)}]{jing02}
{Jing}, Y.~P., \& {Suto}, Y. 2002, \apj, 574, 538

\bibitem[{{Johnston}(1998)}]{johnston98}
{Johnston}, K.~V. 1998, \apj, 495, 297

\bibitem[{{Kandrup} \& {Mahon}(1994)}]{kandrup94}
{Kandrup}, H.~E., \& {Mahon}, M.~E. 1994, \aap, 290, 762

\bibitem[{{Kandrup} {et~al.}(2000){Kandrup}, {Pogorelov}, \&
  {Sideris}}]{kandrup00}
{Kandrup}, H.~E., {Pogorelov}, I.~V., \& {Sideris}, I.~V. 2000, \mnras, 311,
  719

\bibitem[{{Kandrup} \& {Siopis}(2003)}]{kandrup03}
{Kandrup}, H.~E., \& {Siopis}, C. 2003, \mnras, 345, 727

\bibitem[{{Kazantzidis} {et~al.}(2004){Kazantzidis}, {Kravtsov}, {Zentner},
  {Allgood}, {Nagai}, \& {Moore}}]{kazantzidis04}
{Kazantzidis}, S., {Kravtsov}, A.~V., {Zentner}, A.~R., {et~al.} 2004, \apjl,
  611, L73

\bibitem[{{K{\"u}pper} {et~al.}(2015){K{\"u}pper}, {Balbinot}, {Bonaca},
  {Johnston}, {Hogg}, {Kroupa}, \& {Santiago}}]{kuepper15}
{K{\"u}pper}, A.~H.~W., {Balbinot}, E., {Bonaca}, A., {et~al.} 2015, \apj, 803,
  80

\bibitem[{{K{\"u}pper} {et~al.}(2012){K{\"u}pper}, {Lane}, \&
  {Heggie}}]{kuepper12}
{K{\"u}pper}, A.~H.~W., {Lane}, R.~R., \& {Heggie}, D.~C. 2012, \mnras, 420,
  2700

\bibitem[{{Kuzmin}(1973)}]{kuzmin73}
{Kuzmin}, G.~G. 1973, in Dynamics of Galaxies and Star Clusters, ed. T.~B.
  {Omarov}, 71--75

\bibitem[{{Laskar}(1988)}]{laskar88}
{Laskar}, J. 1988, \aap, 198, 341

\bibitem[{{Laskar}(1993)}]{laskar93}
---. 1993, Celestial Mechanics and Dynamical Astronomy, 56, 191

\bibitem[{{Laskar}(1996)}]{laskar96}
---. 1996, Celestial Mechanics and Dynamical Astronomy, 64, 115

\bibitem[{Laskar(1999)}]{laskar99}
Laskar, J. 1999, in NATO ASI Series, Vol. 533, Hamiltonian Systems with Three
  or More Degrees of Freedom (Springer Netherlands), 134--150

\bibitem[{{Laskar}(2003)}]{laskar03}
{Laskar}, J. 2003, ArXiv Mathematics e-prints, math/0305364

\bibitem[{{Laskar} \& {Robutel}(1993)}]{laskar93b}
{Laskar}, J., \& {Robutel}, P. 1993, \nat, 361, 608

\bibitem[{{Law} \& {Majewski}(2010)}]{law10}
{Law}, D.~R., \& {Majewski}, S.~R. 2010, \apj, 714, 229

\bibitem[{{Lee} \& {Suto}(2003)}]{leesuto03}
{Lee}, J., \& {Suto}, Y. 2003, \apj, 585, 151

\bibitem[{{Lichtenberg} \& {Lieberman}(1983)}]{lichtenberg83}
{Lichtenberg}, A.~J., \& {Lieberman}, M.~A. 1983, {Regular and stochastic
  motion}

\bibitem[{Lyapunov(1992)}]{lyapunov92}
Lyapunov, A.~M. 1992, International Journal of Control, 55, 531

\bibitem[{{Maffione} {et~al.}(2015){Maffione}, {G{\'o}mez}, {Cincotta},
  {Giordano}, {Cooper}, \& {O'Shea}}]{maffione15}
{Maffione}, N.~P., {G{\'o}mez}, F.~A., {Cincotta}, P.~M., {et~al.} 2015,
  \mnras, 453, 2830

\bibitem[{{Merritt} \& {Valluri}(1996)}]{merritt96}
{Merritt}, D., \& {Valluri}, M. 1996, \apj, 471, 82

\bibitem[{{Merritt} \& {Valluri}(1999)}]{merritt99}
---. 1999, \aj, 118, 1177

\bibitem[{{Moore} {et~al.}(1998){Moore}, {Governato}, {Quinn}, {Stadel}, \&
  {Lake}}]{moore98}
{Moore}, B., {Governato}, F., {Quinn}, T., {Stadel}, J., \& {Lake}, G. 1998,
  \apjl, 499, L5

\bibitem[{{Navarro} {et~al.}(1996){Navarro}, {Frenk}, \& {White}}]{navarro96}
{Navarro}, J.~F., {Frenk}, C.~S., \& {White}, S.~D.~M. 1996, \apj, 462, 563

\bibitem[{{Ngan} {et~al.}(2015){Ngan}, {Bozek}, {Carlberg}, {Wyse}, {Szalay},
  \& {Madau}}]{ngan15}
{Ngan}, W., {Bozek}, B., {Carlberg}, R.~G., {et~al.} 2015, \apj, 803, 75

\bibitem[{{Odenkirchen} {et~al.}(2001){Odenkirchen}, {Grebel}, {Rockosi},
  {Dehnen}, {Ibata}, {Rix}, {Stolte}, {Wolf}, {Anderson}, {Bahcall},
  {Brinkmann}, {Csabai}, {Hennessy}, {Hindsley}, {Ivezi{\'c}}, {Lupton},
  {Munn}, {Pier}, {Stoughton}, \& {York}}]{odenkirchen01}
{Odenkirchen}, M., {Grebel}, E.~K., {Rockosi}, C.~M., {et~al.} 2001, \apjl,
  548, L165

\bibitem[{{Papaphilippou} \& {Laskar}(1996)}]{papaphilippou96}
{Papaphilippou}, Y., \& {Laskar}, J. 1996, \aap, 307, 427

\bibitem[{{Papaphilippou} \& {Laskar}(1998)}]{papaphilippou98}
---. 1998, \aap, 329, 451

\bibitem[{{Pearson} {et~al.}(2015){Pearson}, {K{\"u}pper}, {Johnston}, \&
  {Price-Whelan}}]{pearson15}
{Pearson}, S., {K{\"u}pper}, A.~H.~W., {Johnston}, K.~V., \& {Price-Whelan},
  A.~M. 2015, \apj, 799, 28

\bibitem[{Pedregosa {et~al.}(2011)Pedregosa, Varoquaux, Gramfort, Michel,
  Thirion, Grisel, Blondel, Prettenhofer, Weiss, Dubourg, Vanderplas, Passos,
  Cournapeau, Brucher, Perrot, \& Duchesnay}]{scikitlearn}
Pedregosa, F., Varoquaux, G., Gramfort, A., {et~al.} 2011, Journal of Machine
  Learning Research

\bibitem[{Price-Whelan(2015)}]{superfreq}
Price-Whelan, A.~M. 2015, SuperFreq, doi:10.5281/zenodo.18787

\bibitem[{{Price-Whelan} {et~al.}(2014){Price-Whelan}, {Hogg}, {Johnston}, \&
  {Hendel}}]{apw14}
{Price-Whelan}, A.~M., {Hogg}, D.~W., {Johnston}, K.~V., \& {Hendel}, D. 2014,
  \apj, 794, 4

\bibitem[{{Price-Whelan} \& {Johnston}(2013)}]{apw13}
{Price-Whelan}, A.~M., \& {Johnston}, K.~V. 2013, \apjl, 778, L12

\bibitem[{Prince \& Dormand(1981)}]{prince81}
Prince, P., \& Dormand, J. 1981, Journal of Computational and Applied
  Mathematics, 7, 67

\bibitem[{{Romanowsky} \& {Kochanek}(1998)}]{romanowsky98}
{Romanowsky}, A.~J., \& {Kochanek}, C.~S. 1998, \apj, 493, 641

\bibitem[{{Sanders}(2014)}]{sanders14}
{Sanders}, J.~L. 2014, \mnras, 443, 423

\bibitem[{{Sanders} \& {Binney}(2013)}]{sanders13a}
{Sanders}, J.~L., \& {Binney}, J. 2013, \mnras, 433, 1813

\bibitem[{{Sesar} {et~al.}(2015){Sesar}, {Bovy}, {Bernard}, {Caldwell},
  {Cohen}, {Fouesneau}, {Johnson}, {Ness}, {Ferguson}, {Martin},
  {Price-Whelan}, {Rix}, {Schlafly}, {Burgett}, {Chambers}, {Flewelling},
  {Hodapp}, {Kaiser}, {Magnier}, {Platais}, {Tonry}, {Waters}, \&
  {Wyse}}]{sesar15a}
{Sesar}, B., {Bovy}, J., {Bernard}, E.~J., {et~al.} 2015, \apj, 809, 59

\bibitem[{{Siegal-Gaskins} \& {Valluri}(2008)}]{siegalgaskins08}
{Siegal-Gaskins}, J.~M., \& {Valluri}, M. 2008, \apj, 681, 40

\bibitem[{Tabor(1989)}]{tabor89}
Tabor, M. 1989, {Chaos and integrability in nonlinear dynamics: an
  introduction} (New York, NY: Wiley)

\bibitem[{Towns {et~al.}(2014)Towns, Cockerill, Dahan, Foster, Gaither,
  Grimshaw, Hazlewood, Lathrop, Lifka, Peterson, Roskies, Scott, \&
  Wilkens-Diehr}]{xsede}
Towns, J., Cockerill, T., Dahan, M., {et~al.} 2014, Computing in Science and
  Engineering, 16, 62

\bibitem[{{Valluri} {et~al.}(2010){Valluri}, {Debattista}, {Quinn}, \&
  {Moore}}]{valluri10}
{Valluri}, M., {Debattista}, V.~P., {Quinn}, T., \& {Moore}, B. 2010, \mnras,
  403, 525

\bibitem[{{Valluri} {et~al.}(2012){Valluri}, {Debattista}, {Quinn}, {Ro{\v
  s}kar}, \& {Wadsley}}]{valluri12}
{Valluri}, M., {Debattista}, V.~P., {Quinn}, T.~R., {Ro{\v s}kar}, R., \&
  {Wadsley}, J. 2012, \mnras, 419, 1951

\bibitem[{{Valluri} {et~al.}(2013){Valluri}, {Debattista}, {Stinson}, {Bailin},
  {Quinn}, {Couchman}, \& {Wadsley}}]{valluri13}
{Valluri}, M., {Debattista}, V.~P., {Stinson}, G.~S., {et~al.} 2013, \apj, 767,
  93

\bibitem[{{Valluri} \& {Merritt}(1998)}]{valluri98}
{Valluri}, M., \& {Merritt}, D. 1998, \apj, 506, 686

\bibitem[{{van Uitert} {et~al.}(2012){van Uitert}, {Hoekstra}, {Schrabback},
  {Gilbank}, {Gladders}, \& {Yee}}]{vanuitert12}
{van Uitert}, E., {Hoekstra}, H., {Schrabback}, T., {et~al.} 2012, \aap, 545,
  A71

\bibitem[{{Varghese} {et~al.}(2011){Varghese}, {Ibata}, \&
  {Lewis}}]{varghese11}
{Varghese}, A., {Ibata}, R., \& {Lewis}, G.~F. 2011, \mnras, 417, 198

\bibitem[{{Vera-Ciro} {et~al.}(2011){Vera-Ciro}, {Sales}, {Helmi}, {Frenk},
  {Navarro}, {Springel}, {Vogelsberger}, \& {White}}]{veraciro11}
{Vera-Ciro}, C.~A., {Sales}, L.~V., {Helmi}, A., {et~al.} 2011, \mnras, 416,
  1377

\bibitem[{{Vogelsberger} {et~al.}(2008){Vogelsberger}, {White}, {Helmi}, \&
  {Springel}}]{vogelsberger08}
{Vogelsberger}, M., {White}, S.~D.~M., {Helmi}, A., \& {Springel}, V. 2008,
  \mnras, 385, 236

\bibitem[{{Wechsler} {et~al.}(2002){Wechsler}, {Bullock}, {Primack},
  {Kravtsov}, \& {Dekel}}]{wechsler02}
{Wechsler}, R.~H., {Bullock}, J.~S., {Primack}, J.~R., {Kravtsov}, A.~V., \&
  {Dekel}, A. 2002, \apj, 568, 52

\bibitem[{{Wegg} \& {Gerhard}(2013)}]{wegg13}
{Wegg}, C., \& {Gerhard}, O. 2013, \mnras, 435, 1874

\bibitem[{{Wegg} {et~al.}(2015){Wegg}, {Gerhard}, \& {Portail}}]{wegg15}
{Wegg}, C., {Gerhard}, O., \& {Portail}, M. 2015, \mnras, 450, 4050

\bibitem[{Wolf {et~al.}(1985)Wolf, Swift, Swinney, \& Vastano}]{wolf85}
Wolf, A., Swift, J.~B., Swinney, H.~L., \& Vastano, J.~A. 1985, Physica, 285

\bibitem[{{Zemp} {et~al.}(2009){Zemp}, {Diemand}, {Kuhlen}, {Madau}, {Moore},
  {Potter}, {Stadel}, \& {Widrow}}]{zemp09}
{Zemp}, M., {Diemand}, J., {Kuhlen}, M., {et~al.} 2009, \mnras, 394, 641

\bibitem[{{Zemp} {et~al.}(2011){Zemp}, {Gnedin}, {Gnedin}, \&
  {Kravtsov}}]{zemp11}
{Zemp}, M., {Gnedin}, O.~Y., {Gnedin}, N.~Y., \& {Kravtsov}, A.~V. 2011, \apjs,
  197, 30

\bibitem[{{Zhu} {et~al.}(2015){Zhu}, {Marinacci}, {Maji}, {Li}, {Springel}, \&
  {Hernquist}}]{zhu15}
{Zhu}, Q., {Marinacci}, F., {Maji}, M., {et~al.} 2015, ArXiv e-prints,
  arXiv:1506.05537

\end{thebibliography}

\appendix
\section{Lyapunov exponents} \label{sec:lyapapdx}

%Using the definition of $w$ from Equation~\ref{eq:coords}, we can write Hamilton's equations as \footnote{A sum is implied with any repeated indices.}
%\begin{equation}
%	\dot{w}_i = \mathcal{J}_{ik}\,\frac{\partial H}{\partial w_k} \label{eq:ham}
%\end{equation}
%where $\mathcal{J}_{ik}$ is the $2N \times 2N$ canonical Poisson tensor (also called the symplectic matrix) defined by
%\begin{equation}
%	\mathcal{J}_{ik} = \left( \begin{array}{c:c} 0 & \ident \\ \hdashline -\ident & 0 \end{array} \right)
%\end{equation}
%with $N$-dimensional identity matrices $\ident$. We will consider a nearby phase-space position, $w_i'$, separated from $w_i$ by an infinitesimal deviation, $\delta w_i$, such that $w_i' = w_i + \delta w_i$. We can expand to linear order in the deviation about the parent orbit and write the equations of motion for the deviation as
%\begin{align}
%	\dot{\delta w_i} &= \mathcal{J}_{ik} \, \frac{\partial^2 H}{\partial w_k \partial w_m} \, \delta w_m =  \mathcal{J}_{ik}\, D_{km} \, \delta w_m \\
%	&= A_{im} \, \delta w_m
%\end{align}
%where $D_{km}$ is the Hessian matrix evaluated at the parent orbit. The general solution for this equation is
%\begin{equation}
%	\delta w_i(t) = B_{im}(t) \, \delta w_i(0)
%\end{equation}
%where $\delta w_i(0)$ are the initial conditions for the deviation vector and $B_{im}$ is the solution matrix.
% For chaotic orbits, the maximum eigenvalue of the solution matrix to Eq.~\ref{eq:deviate} is positive real, leading to exponential divergence of nearby orbits.

If one is only interested in characterizing the degree of chaos, computing the full Lyapunov spectrum for an orbit is often not necessary. It is usually sufficient to compute an estimate of the maximum Lyapunov exponent by estimating the finite-time maximum Lyapunov exponent (FTMLE). Using the definition of $\boldsymbol{w}$ from Equation~\ref{eq:coords}, consider an orbit that is a small deviation away from the parent orbit, $\boldsymbol{w}' = \boldsymbol{w} + \delta\boldsymbol{w}$. If the parent orbit is chaotic, the norm of the infinitesimal deviation should grow exponentially with time with some characteristic rate, $\lambda$,
\begin{equation}
	\|\delta\boldsymbol{w}(t)\| = e^{\lambda \, t} \, \|\delta\boldsymbol{w}_0\|
\end{equation}
\citep[see, e.g.,][]{lichtenberg83,tabor89}. From this expression, we see that
\begin{equation}
	\lambda(t) = \frac{1}{t}\ln \frac{\|\delta \bs{w}(t)\|}{\|\delta \bs{w}_0\|} \label{eq:mle}
\end{equation}
where the maximum Lyapunov exponent is the limit as $t\rightarrow \infty$,
\begin{equation}
	\lambda_{\rm max} = \lim_{t\rightarrow\infty}\lambda(t). \label{eq:lmax}
\end{equation}
Numerically computing this quantity is not trivial because (1) obviously the limit to infinity is not possible and (2) the norm of the deviation vector $\|\delta \bs{w}(t)\|$ is expected to increase exponentially for chaotic orbits, leading to nonlinear evolution of the deviation and numerical problems. To circumvent these issues, it is sufficient to instead start a nearby orbit with some small initial deviation with norm $\delta_0$, integrate for a sufficiently small amount of time, $\tau$, then renormalize the deviation back to the initial norm \citep{benettin76}. There is no general way to determine $\tau$ except to perform convergence tests.
%The following pseudocode outlines this procedure:\footnote{see Gary for Python implementation?}\\
%\begin{algorithmic}[1]
%\State {\bf define} orbital integration timestep, $h$, and number of steps, $K$
%\State {\bf define} initial norm of deviation vector, $\delta_0$, to be sufficiently small
%\State {\bf define} renormalization integration period, $\tau \ll hK$
%\For{each timestep when integrating the main orbit, $\bs{w}(t)$}
%\State step forward the orbit and deviation vector orbit by one timestep, $t_{i-1} \rightarrow t_i$
%\If {a normalization timestep}
%\State measure and store the length of the deviation vector, $\delta_i = \|\delta \bs{w}_i\|$
%\State renormalize the length of the deviation vector, $\delta \bs{w}_i = \delta \bs{w}_i (\delta_0/\delta_i)$
%\EndIf
%\EndFor
%\end{algorithmic}
The FTMLE after a given number of timesteps, $N$, is then estimated as
\begin{equation}
	\lambda_N = \frac{1}{t_N}\sum_i^N \ln \frac{\|\delta \bs{w}(t_i)\|}{\|\delta \bs{w}_0\|} \label{eq:ftmle}
\end{equation}
where the $t_i$ are the times at which the renormalization occurs and the MLE is estimated after a very long time to approximate the limit of Equation~\ref{eq:lmax}.

For most regular orbits, deviations will grow linearly or as a power-law of time. As we have seen in Section~\ref{sec:nldreview}, if the orbit is regular, there exists a local transformation to action-angle variables where the angle variables increase linearly with time, $\theta_i \propto \Omega_i t$. We can look at small variations around the angle-space orbit,
\begin{align}
	\frac{d (\delta \theta_i)}{dt} &= \frac{\partial \Omega_i}{\partial J_k} \delta J_k
\end{align}
These equations are easily integrated:
\begin{align}
	\delta \theta_i(t) &= \delta \theta_i(0) + \left(\frac{\partial \Omega_i}{\partial J_k} \delta J_k \right) t
\end{align}
It's clear then that the norm of the deviation vector grows linearly with time:
\begin{align}
	\|\delta \bs{w}(t)\| &= \left[\sum_i (\delta \theta_i(t))^2 + \sum_i (\delta J_i)^2\right]^{1/2}\\
	&= \left[\sum_i \left(\delta \theta_i(0) + \frac{\partial \Omega_i}{\partial J_k} \delta J_k t\right)^2 + \sum_i (\delta J_i)^2\right]^{1/2}\\
	&\propto t.
\end{align}
From Equations~\ref{eq:mle} and \ref{eq:lmax} it is evident that any deviation vector that grows as a power law with time, $t^k$, will asymptote to 0 from the limit
\begin{equation}
	\lambda_{\rm max} \propto \lim_{t\rightarrow \infty} k \frac{\ln t}{t} = 0.
\end{equation}
At long times, the numerically computed MLE for a regular orbit should approach 0 as $t^{-1}$. For chaotic orbits, the divergence is exponential, and the limit should converge to the rate of the exponential: the Lyapunov exponent. In practice, the MLE is often estimated as the mean of $\lambda_N$ after the summation diverges from power-law behavior. For weakly chaotic orbits, reliable computation of the MLE may take integration of thousands of orbital periods.

\section{SuperFreq}\label{sec:naffapdx}

\renewcommand{\thefigure}{\thesection.\arabic{figure}}

The NAFF algorithm operates on numerically integrated orbital time-series (i.e. positions and velocities at a set of equally spaced times). Complex combinations of the phase-space coordinates of the orbit are Fourier-transformed using a standard FFT and the resultant spectrum is then searched for the strongest frequency component; this serves as an initial guess for the frequency of a particular Fourier component. Solving for the frequency that maximizes the Fourier integral of the FFT with a particular window function allows for a more accurate determination of the frequency. It has been shown that this accuracy converges much faster than the typical $\inttime^{-1}$ expected for a standard FFT by using a window function of the form
\begin{equation}
	W(\tau=t/\inttime) = \frac{2^p \, (p!)^2}{(2p)!} \left( 1 + \cos \pi \tau\right)^p
\end{equation}
where $\inttime$ is the integration time \citep{laskar99}. Once the strongest frequency component is found, it is subtracted from the spectrum and the process is repeated iteratively. This procedure generates a table of frequencies for each component of motion which must then be searched for the three fundamental frequencies. The original implementations of NAFF have used $p=1$ \citep[e.g.,][]{laskar93, valluri98}, but we have chosen to use $p=4$. With a higher order window function, the central peak is broadened, but the amplitudes of the side lobes decrease faster \citep[see][]{hunter02}; we have found that this additional damping of the side lobes allows for more reliable determination of frequencies from strongly chaotic orbits.

\setcounter{figure}{0}
\begin{figure}
\begin{center}
\includegraphics[width=\textwidth]{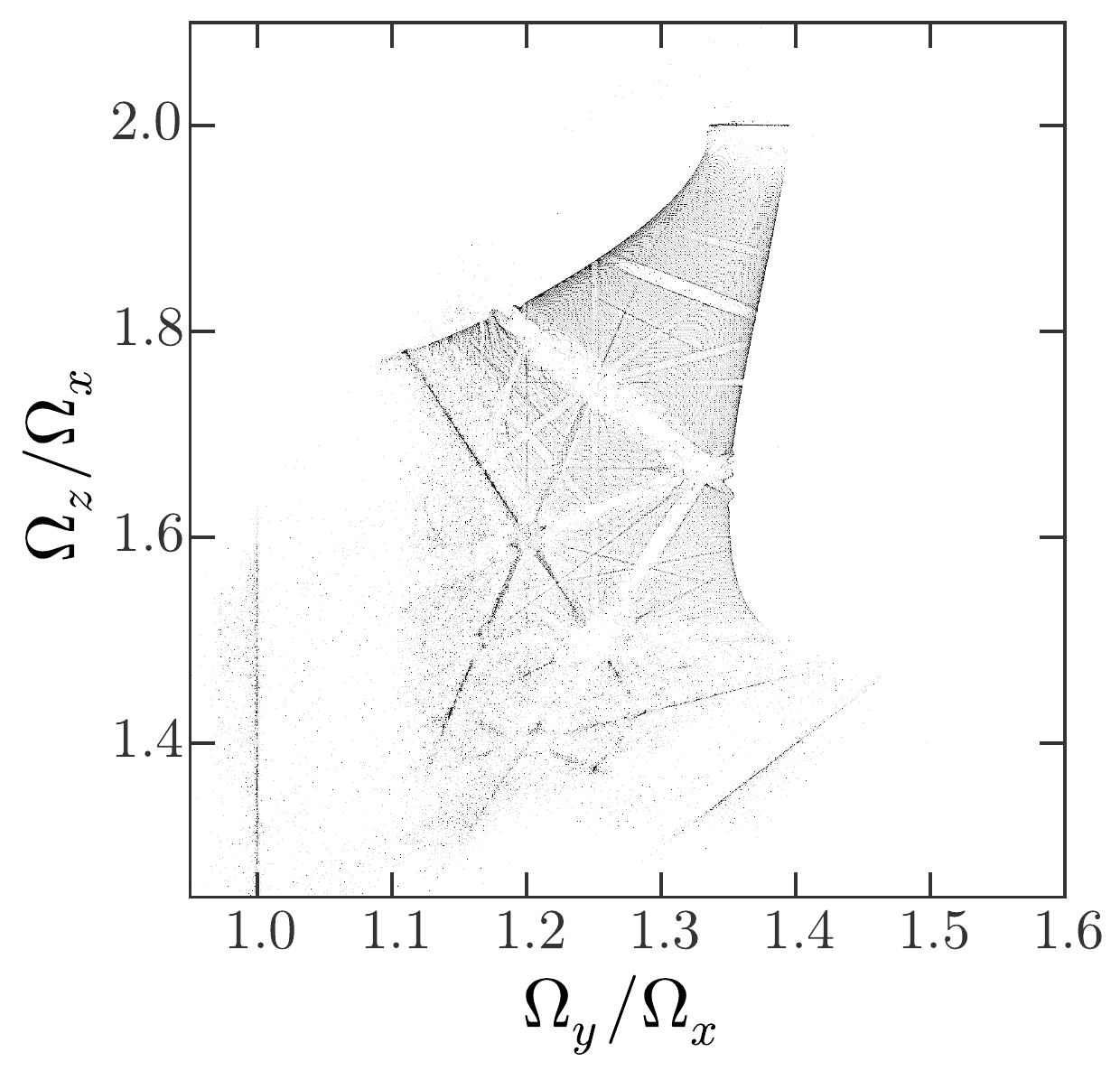}
\caption{ A reproduction of Figure~3.45 from \citep{binneytremaine} as a validation of our frequency analysis code: Frequency ratios for 100000 isoenergy orbits integrated in a triaxial logarithmic potential. Linear features are resonances---stable resonances appear as dark lines, unstable resonances appear as linear gaps.} \label{fig:logfreqs}
\end{center}
\end{figure}

\superfreq\ recovers the fundamental frequencies for an orbit faster (with a fewer number of terms) when the coordinates used are `close' to the angle variables \cite[PL96;][]{papaphilippou96}. PL96 show that a good choice of coordinates for tube orbits are the Poincar\'e symplectic polar coordinates, a set of canonical coordinates similar to cylindrical coordinates. When computing the frequencies for tube orbits, we first align the circulation about the $z$-axis through rotation, transform to Poincar\'e polar coordinates, then use \superfreq\ to measure the fundamental frequencies. We could equivalently use the Cartesian time series, but the convergence of terms is slower (the amplitudes of successive terms decrease slower for Cartesian coordinates). We have tested that our implementation of \superfreq\ returns the same fundamental frequencies in either case for a set of tube orbits. For box orbits, the motion is close to separable in each Cartesian component and we therefore use Cartesian coordinates for estimating the frequencies for these orbits.

Figure~\ref{fig:logfreqs} shows a validation of our frequency analysis code in which we reproduce the frequency map at a fixed energy of an axisymmetric, logarithmic potential \cite[][pg. 260, Figure~3.45]{binneytremaine}. Plotted are the (Cartesian) frequency ratios recovered for a grid of iso-energy, box orbits integrated in the potential
\begin{equation}
	\Phi(x,y,z) = \frac{1}{2}\ln\left(x^2 + (y/0.9)^2 + (z/0.7)^2 + 0.1\right). \label{eq:logpotential}
\end{equation}
Following \cite{binneytremaine}, we generate a grid of orbits on the equipotential surface $\Phi(x,y,z) = 0.5$ (we use a grid with 100000 orbits compared to their 10000 orbits). Each orbit is integrated for $\approx$40 orbital periods. In this figure, stable resonances appear as linear over-densities and unstable resonances appear as linear under-densities or gaps. The regularity of the points in this map reflects the input grid of initial conditions. Points that appear to be erratically scattered are chaotic orbits where the frequencies change with time.

% ----------------------------------------------------------------------------------

\renewcommand{\thefigure}{\arabic{figure}}

% Figure ??
%\begin{figure}[!p]
%\begin{center}
%\includegraphics[width=\textwidth]{figures/potential.pdf}
%\caption{Equipotential contours for the triaxial NFW potential considered in this work. There are eight contour levels evenly spaced and linear in the value of the potential. } \label{fig:potential}
%\end{center}
%\end{figure}

\end{document}